%% file: main.tex
\documentclass[journal=jacsat,manuscript=article]{achemso}

\input{preamble.tex}

\graphicspath{illustrations/}
\input{abbreviations}

\author{Jonas Böhm}
\affiliation[CNRS]{Université de Bordeaux, CNRS, Bordeaux INP, ICMCB, F-33600 Pessac, France}
\affiliation[RS2E]{Réseau sur le Stockage Electrochimique de l’Energie (RS2E), CNRS FR 3459, Cedex 1, F-80039 Amiens, France}
\email{jonas.bohm@icmcb.cnrs.fr}
\author{Aurélie Champagne}
\affiliation[CNRS]{Université de Bordeaux, CNRS, Bordeaux INP, ICMCB, F-33600 Pessac, France}
\affiliation[RS2E]{Réseau sur le Stockage Electrochimique de l’Energie (RS2E), CNRS FR 3459, Cedex 1, F-80039 Amiens, France}
\email{aurelie.champagne@icmcb.cnrs.fr}

\title{Predicting Crystal Structures and Ionic Conductivities in Li$\lo{3}$YCl$\lo{6-x}$Br$\lo{x}$ Halide Solid Electrolytes Using a Fine-Tuned Machine Learning Interatomic Potential}

\begin{document}

\begin{abstract}
Understanding ionic transport in halide solid electrolytes is essential for advancing next-generation solid-state batteries. This work demonstrates the effectiveness of fine-tuning the Crystal Hamiltonian Graph Network (CHGNet) universal machine learning interatomic potential to accurately predict total energies, relaxed geometries, and lithium-ion dynamics in the ternary halide family Li$\lo{3}$YCl$\lo{6-x}$Br$\lo{x}$. Starting from experimentally refined disordered structures of Li$\lo{3}$YCl$\lo{6}$
and Li$\lo{3}$YBr$\lo{6}$, we present a strategy for generating ordered structural models through systematic enumeration and energy ranking, providing realistic structural models. These serve as initial configurations for an iterative fine-tuning workflow that integrates molecular dynamics simulations and static density functional theory calculations to achieve near-ab initio accuracy at four orders of magnitude lower computational cost. We further reveal the influence of composition (varied x) on the predicted phase stability and ionic conductivity in Li$\lo{3}$YCl$\lo{6-x}$Br$\lo{x}$, demonstrating the robustness of our approach for modeling transport properties in complex solid electrolytes.
\end{abstract}


\section{Introduction}

Lithium-ion (Li-ion) batteries play a crucial role in modern energy storage, powering consumer electronics, electric vehicles, and grid-level applications thanks to their high gravimetric energy density and long life cycle. They not only enable the transition toward sustainable mobility but also help stabilize power grids by compensating for the intermittency of renewable energy sources~\cite{daniel_materials_2008,goodenough_li-ion_2013}. A conventional Li-ion battery consists of a negative electrode, liquid electrolyte, separator, and positive electrode. In contrast, \gls{assbs}, replace the flammable liquid electrolyte and polymer separator with a \gls{se}, offering enhanced safety, higher energy and power density, faster charging~\cite{ohno_materials_2020}, and reduced cross-contamination between electrodes~\cite{janek_challenges_2023}. However, realizing these advantages introduces major materials and engineering challenges~\cite{janek_challenges_2023}, including: (1) developing solid composite cathodes that operate reliably under minimal pressure ($< 0.1$\,MPa); (2) identifying \glspl{se} with  high ionic conductivity, broad electrochemical stability window, mechanical robustness, and low cost; (3) ensuring stable, low-resistance interfaces between \glspl{se} and electrodes; and (4) implementing high-capacity anodes such as Li metal or Si-based materials for enhanced battery energy density. Among these, the \gls{se} is central, as it governs ion transport, interfacial stability, and mechanical integrity $-$ directly impacting the first three challenges~\cite{janek_challenges_2023,kerman2017review,pasta2020roadmap}. An ideal \gls{se} should combine high ionic conductivity (\qty{\ge 10}{mS.cm}$^{-1}$), negligible electronic conductivity, mechanical strength sufficient to suppress dendrite growth, and (electro)chemical stability across the $0-5$\,V vs. Li/Li$^+$ window~\cite{janek_challenges_2023}.\\

Recently, ternary metal halides with general formula Li$\lo{3}$MA$\lo{6-x}$B$\lo{x}$ (M = Y, Er, and In; A, B = Cl, Br, I) have emerged as highly promising \glspl{se}~\cite{asano_solid_2018}. These materials offer a unique combination of electrochemical stability compatible with 4\,V-class cathodes, mechanical and thermal robustness, and room temperature Li$^+$ conductivities exceeding \qty{1}{mS.cm}$^{-1}$~\cite{asano_solid_2018}. Their ionic conductivity can be further tuned through compositional substitution and synthesis control~\cite{LYB_LYC_exp,asano_solid_2018,van_der_maas_investigation_2023,liu_tuning_2024}, providing a versatile compositional design space for performance optimization. For instance, van der Maas \textit{et al.} observed in the \gls{lycb} series that ionic conductivity peaks at \qty{5.36}{mS.cm}$^{-1}$ near $\text{x}=4.5$~\cite{van_der_maas_investigation_2023}. They also observed a structural transition from trigonal ($P\bar{3}m1$) to monoclinic ($C2/m$) symmetry within $1.5<\text{x}<3$; however the interplay between this structural change and the enhancement in ion transport remains unclear.

Designing halide electrolytes with increased ionic conductivity thus requires a microscopic understanding of how composition, structure, and anion substitution modulate Li$^+$ migration pathways. Yet, the vast compositional space and experimental limitations in directly probing atomic-scale diffusion processes necessitate integrated computational-experimental approaches. Atomistic modeling, especially \gls{md}, provides valuable insights into Li$^+$ diffusion mechanisms and structure-transport relationships, enabling to narrow down the most promising compositions.

The accuracy of \gls{md} simulations relies on the description of the underlying \gls{pes}, traditionally obtained either from first-principles methods such as \gls{dft} or from empirical force fields. While \gls{dft} offers high accuracy, it is computationally prohibitive for the long timescales required to capture diffusion processes (on the order of several nanoseconds). \glspl{mlip} have recently emerged as a powerful alternative, combining near-\gls{dft} accuracy with several orders of magnitude lower computational cost~\cite{wang_machine_2024}. \glspl{mlip} approximate the \gls{pes} by mapping atomic coordinates $\arr{\boldsymbol{r}\lo{v}}$ and atomic numbers $\arr{Z\tief{v}}$ to total energies $E$~\cite{jacobs_practical_2025}, using local environment descriptors and \gls{ml} regression schemes. Their linear scaling with system size and parallelizability make them ideally suited for large-scale \gls{md} simulations~\cite{7net}. The performance of an \gls{mlip} critically depends on its training data; and large datasets are often required to achieve robust generalization. Pretrained \glspl{umlip} $-$ such as \gls{m3gnet}~\cite{m3gnet}, \gls{chgnet}~\cite{chgnet}, and \gls{7net}~\cite{7net} $-$ are trained on extensive materials databases encompassing diverse chemical systems~\cite{mat_proj,mat_proj_traj,barroso_omat24}, and can, in principle, generalize across chemistries. However, their transferability to new material classes such as halide \glspl{se} remains uncertain. Thus, system-specific fine-tuning is often required to achieve reliable quantitative predictions~\cite{liu_study_2025,huang_cross-functional_2025,kaur_data-efficient_2025,radova_fine-tuning_2025}.\\

In this work, we investigate lithium-ion transport in halide \glspl{se} by elucidating how crystal structure, stoichiometry, and anion substitution govern ionic conductivity within the \gls{lycb} family. We first present an efficient and accurate enumeration-ranking procedure to systematically explore all possible ordered configurations starting from disordered experimentally-refined structures. We then develop an iterative fine-tuning procedure, inspired by active-learning loops~\cite{devita_on_the_fly,shapeev_active_learning,bernstein_active_learning,vandermause_on_the_fly}, that integrates targeted \gls{dft} data generation with adaptive model refinement. Using this approach, we fine-tune the pretrained \gls{chgnet} model~\cite{chgnet} to reproduce the \gls{pes} of \gls{lycb} compounds with high-fidelity, enabling nanosecond \gls{md} simulations at a computational cost roughly $10^4$ times lower than \gls{dft}. Finally, we benchmark the fine-tuned \gls{chgnet} model against its pretrained counterpart and the state-of-the-art \gls{7net} potential~\cite{7net} (pretrained on an even larger dataset including out-of-equilibrium structures~\cite{barroso_omat24}). Our results provide fundamental insights into Li$^+$ transport and phase stability across halide compositions, and establishes a general, data-efficient strategy for adapting \gls{umlip} to new material systems $-$ a strategy significantly more effective than training a \gls{ml} model from scratch. Overall, this framework paves the way for accelerated discovery and rational design of next-generation \glspl{se} for \gls{assbs}.

\section{Results and Discussion}

\subsection{Structure enumeration and energy-based ranking}

\begin{figure}[tb!]
\centering
\includegraphics[width=1.0\linewidth]{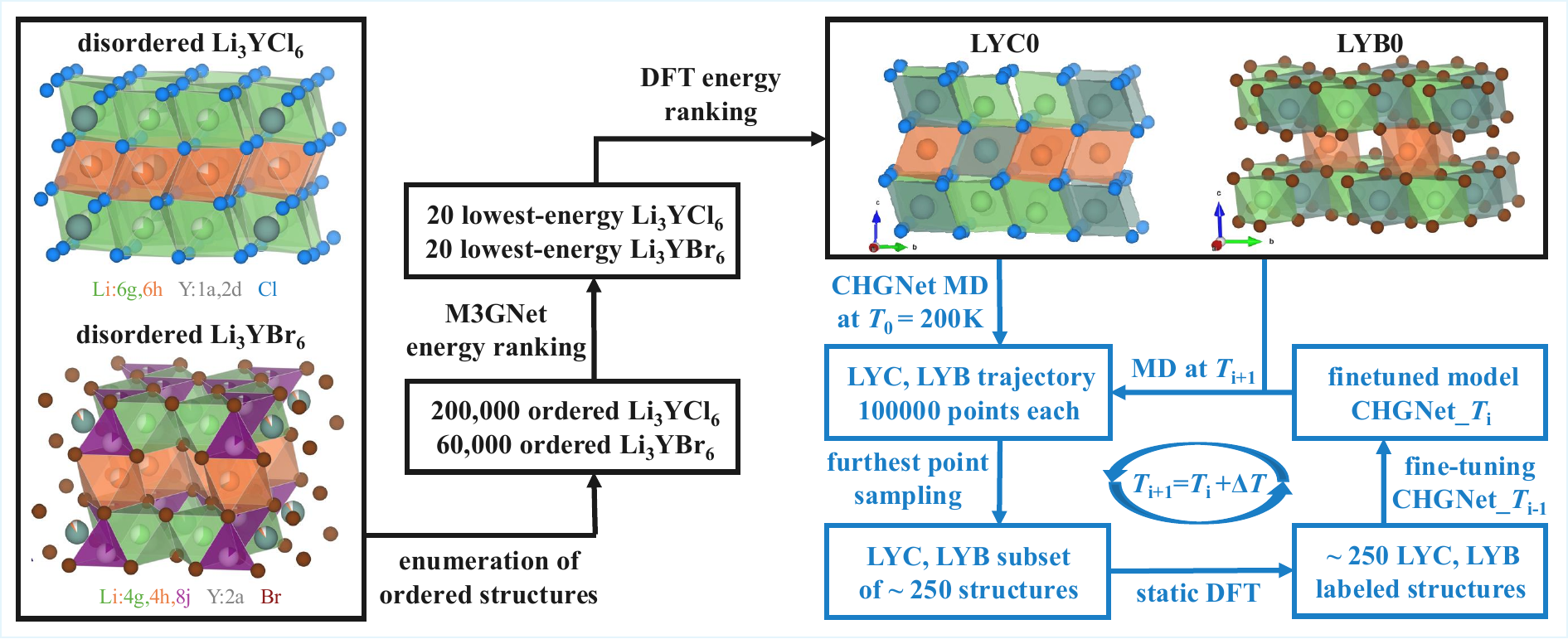}
\caption{Global workflow combining the enumeration-ranking strategy (black boxes) with the \gls{umlip} fine-tuning framework (blue boxes). Starting from experimentally resolved structures~\cite{LYB_LYC_exp} $-$ which exhibit partial occupancies $-$ symmetry-distinct ordered configurations are enumerated and ranked using the \gls{m3gnet} \gls{umlip}. Additional DFT calculations are performed on the lowest-energy configurations to identify the ground-state ordered configuration $-$ subsequently used as a starting configuration in the fine-tuning procedure. The fine-tuning procedure involves a series of \gls{chgnet}-driven \gls{md} simulations (with LAMMPS) at finite temperatures ($T_{\mathrm{i}}$). For each \gls{md} run (1,000,000 steps), DFT calculations (with VASP) are performed on a subset of structures to obtain additional training data that are used to fine-tune \gls{chgnet} model. Computational details are provided in the Computational Methods.}
\label{fig:1-workflow}
\end{figure}

A key initial challenge in modeling halide \glspl{se} lies in defining suitable starting structures. Experimentally refined models of \gls{lyc} ($P\bar{3}m1$) and \gls{lyb} ($C2/m$) exhibit partial occupancies on Li and Y sites, reflecting intrinsic disorder (Figure~\ref{fig:1-workflow}). Determining the lowest-energy ordered configuration among all possible arrangements requires evaluating the total energy of a vast amount of configurations quickly approaching infinity with increasing supercell size. Conventional Ewald-based electrostatic ranking~\cite{ewald_berechnung_1921,toukmaji_ewald_1996} offers a rapid but approximate criterion that neglects many-body interactions, whereas \gls{dft} provides higher accuracy but is computationally prohibitive for such vast configurational spaces. This configurational complexity is not unique to halide conductors and also arises in disordered oxide and sulfide electrodes and electrolytes. Consequently, many of these compounds are absent from large materials databases such as the Materials Project~\cite{mat_proj}, since their accurate representation demands large supercells and explicit sampling of disorder.\\

To overcome these limitations, we employ a \gls{umlip} $-$ namely \gls{m3gnet} model~\cite{m3gnet} $-$ to efficiently screen and rank candidate configurations. Starting from $1\times1\times2$ supercells of the experimental \gls{lyc} and \gls{lyb} models~\cite{LYB_LYC_exp}, all symmetrically distinct orderings are generated using the \textit{enumlib} library~\cite{enum} (Figure~\ref{fig:1-workflow}). A total of $\sim 60,000$ LYC and $\sim 200,000$ LYB structures are optimized using \gls{m3gnet} and ranked according to their predicted total energy. From these, we identify the 20 lowest-energy structures and perform subsequent \gls{dft} optimization to validate their energetic ordering. This procedure yields low-energy ordered structures for both \gls{lyc} and \gls{lyb}, denoted LYC0 ($P\bar{3}c1$) and LYB0 ($C2$), respectively $-$ and more importantly provides a physical starting point for our subsequent \gls{md} simulations.

We extend this approach to mixed-halide compositions, \gls{lycb}, by systematically substituting Cl/Br in both parent lattices, followed by enumeration and \gls{m3gnet}-based optimization and energy ranking. Across the full composition range ($\text{x}=0$ to 6, step 0.5), we obtain 13 ordered derivatives for each parent phase. These ordered models provide consistent, low-energy starting points, which are subsequently used in the \glspl{umlip} fine-tuning procedure and for lithium-ion diffusion simulations.

\subsection{CHGNet performance for energy and volume prediction}

To assess baseline performance, we benchmark the pretrained \gls{umlip} \gls{chgnet} against \gls{dft} reference data.

\begin{figure}[tb!]
\centering
\begin{overpic}[width=1.0\linewidth]{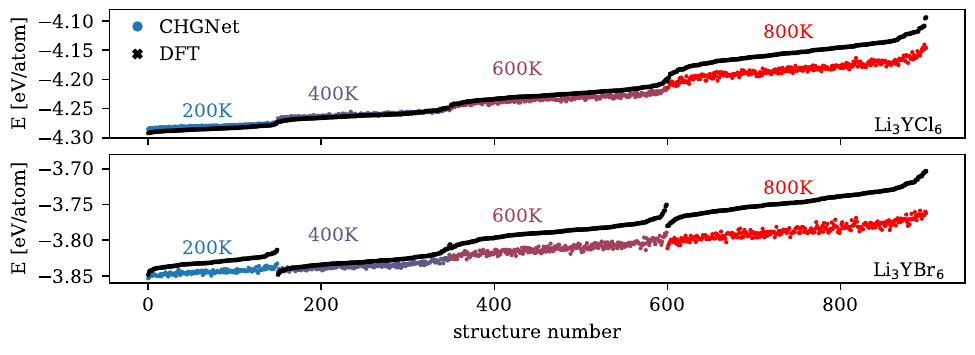}
        \put(10,37){a)} 
\end{overpic}
\begin{overpic}[width=1.0\linewidth]{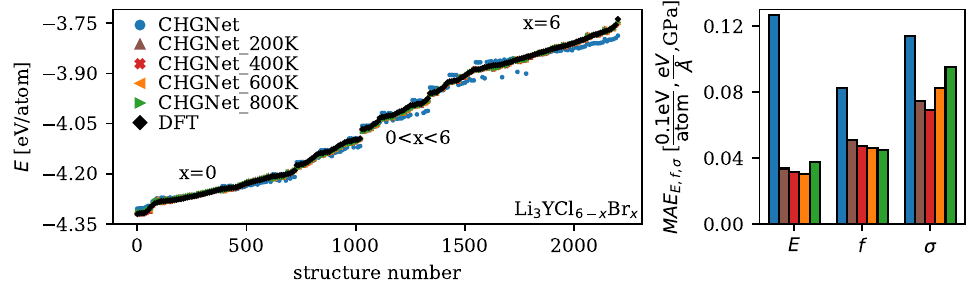}
        \put(10,31){b)} 
        \put(76,31){c)} 
\end{overpic}
\begin{overpic}[width=1.0\linewidth]{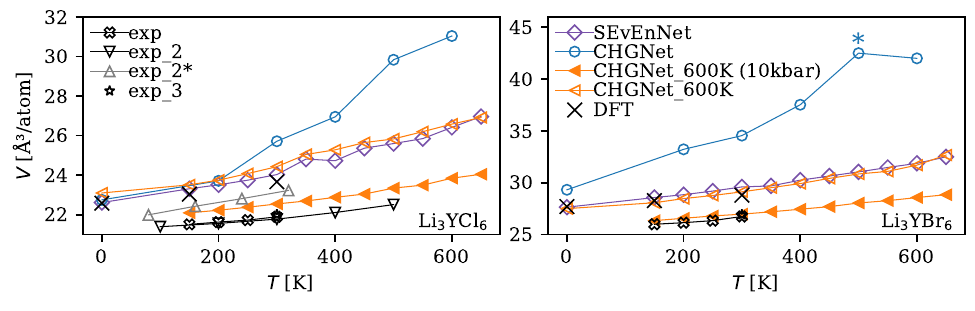}
        \put(9.5,32){d)} 
        \put(56,32){e)} 
\end{overpic}
\caption{(a) Total energies predicted with \gls{chgnet} (colored circles), compared to DFT reference data (black crosses), for a series of finite-temperature structures generated during the fine-tuning procedure in Figure~\ref{fig:1-workflow}. (b) Benchmark of the pretrained and fine-tuned \gls{chgnet} potentials for total energy prediction of \gls{lyc}, \gls{lyb}, and \gls{lycb}, compared to DFT data. (c) MAEs on energy (in 0.1\,eV/atom), forces (in eV/\AA), and stress (in GPa) for the various models with the same color-code as in (b). (d) and (e) Average volume/atom resulting from the $NpT$ equilibration \gls{md} runs, using the pretrained \gls{chgnet} (blue circles) and fine-tuned \gls{chgnet}\_600K potentials (orange triangles, solid for high pressure of 10kbar), compared to \gls{7net} predictions (purple diamonds), DFT calculations (black crosses), and experimental data (black and gray markers) from Refs.~\citenum{LYB_LYC_exp,liu_tuning_2024,asano_solid_2018}. Note the blue star in (e) indicates an unstable simulation with nonphysical cell volume, which is scaled here by 0.1.}
\label{fig:2-benchmark}
\end{figure}

For 0\,K-structures, static \gls{chgnet} calculations reproduce energetic trends of \gls{lycb} compounds with mean absolute errors (MAEs) below \qty{11}{meV.atom}$^{-1}$ (Figure~S1).
Furthermore, \gls{chgnet} structure optimizations predict volumes and cell parameters within 5\% of the \gls{dft} values (Figure~S7), while being over $2,000$ times faster to compute ($\approx 50$\,s CPU vs. $\approx 33$\,h CPU per structure using \gls{dft}). CHGNet's accuracy for energy and structure predictions at finite temperature is benchmarked in Figure~\ref{fig:2-benchmark}. In Figure~\ref{fig:2-benchmark}(a), we compare total energies predicted by \gls{chgnet} (circles) with DFT references (black crosses) for 880 \gls{lyc} and 880 \gls{lyb} structures extracted from finite-temperature \gls{md} simulations. While CHGNet performs remarkably well at low temperature, energy predictions start to deviate at $T\ge\qty{600}{K}$, with errors in the energy predictions as large as \qty{70}{meV.atom}$^{-1}$.

To assess CHGNet's accuracy in volume prediction, we perform \gls{md} simulations in the $NpT$ ensemble and obtain equilibrated structures for \gls{lyc} and \gls{lyb} at targeted temperatures up to 650\,K. Figures~\ref{fig:2-benchmark}(d) and (e) compare \gls{chgnet} predicted volumes with \gls{7net} data and available experimental values: exp (neutron diffraction)~\cite{LYB_LYC_exp}, exp\_2 (neutron diffraction), exp\_2* (synchrotron X-ray diffraction on higher crystalline sample)~\cite{liu_tuning_2024}, and exp\_3 (X-ray diffraction)~\cite{asano_solid_2018}. While \gls{chgnet} matches \gls{7net} volumes within $2-8$\% at low temperature, it significantly overestimates volumes at higher temperature. Moreover, above 500\,K, some simulations became unstable, with diverging volume fluctuations (indicated by a star in Figure~\ref{fig:2-benchmark}(e)). These failures highlight \gls{chgnet}'s limitations in capturing thermodynamic and structural diversity encountered at finite temperature, underscoring the need for system-specific fine-tuning.

\subsection{Fine-tuning \gls{chgnet} with finite-temperature data}

To address this, we implement an iterative fine-tuning protocol that augments the training dataset with targeted \gls{dft} data for \gls{lyc} and \gls{lyb} compounds (see Figure~\ref{fig:1-workflow} and Computational Methods). Four progressively refined models $-$ \gls{chgnet}\_200K, \_400K, \_600K, and \_800K $-$ are trained by successively incorporating structures from \gls{md} trajectories at increasing temperature (Figure~\ref{fig:2-benchmark}(a)), starting from the fine-tuned model from the previous iteration. In the SI, we discuss the effect of fine-tuning CHGNet with all accumulated training data at once. The present active-learning-inspired refinement expands the configurational diversity of the training data and significantly improves agreement with \gls{dft} across all metrics, as shown for an independent test-dataset of \gls{lycb} structures from 0\,K to 800\,K in Figures~\ref{fig:2-benchmark}(b) and (c). Remarkably, although \gls{chgnet} is fine-tuned only on the \gls{lyc} and \gls{lyb} endpoints, it retains high predictive accuracy for finite-temperature intermediate compositions. Among all models, \gls{chgnet}\_600K offers a good balance between the metrics, with energy, force, and stress MAEs below \qty{3.5}{meV.atom}$^{-1}$, \qty{50}{meV.\angstrom}$^{-1}$, and \qty{85}{MPa}, respectively $-$ comparable to state-of-the-art MLIPs~\cite{park_sevennet_mf_ompa_2025}. Importantly, \gls{chgnet}\_600K remains stable during long $NpT$ \gls{md} runs up to 800\,K, where the pretrained model failed. Subsequent analyses are therefore performed using \gls{chgnet}\_600K, if not otherwise stated.\\

Fine-tuning improves not only static energies but also thermodynamic observables. Figures~\ref{fig:2-benchmark}(d), (e) and Tables~\ref{tab:1-lyc}, \ref{tab:2-lyb} compare the average volumes predicted by \gls{chgnet}, \gls{chgnet}\_600K, and \gls{7net} for \gls{lyc} and \gls{lyb}. While the pretrained model systematically overestimates the reference volumes from \gls{7net} and DFT by 2-8\% at 0\,K and up to 25-40\% at higher temperature, the fine-tuned \gls{chgnet}\_600K model brings the predictions into near-quantitative agreement with both \gls{7net} and direct \textit{ab initio} calculations (see Computational Methods), with deviations within 2\%. This level of consistency between \gls{chgnet}\_600K and DFT indicates that the fine-tuned model has reached the accuracy ceiling expected for a \gls{mlip} trained on first-principles data. Relative to experiment, \gls{chgnet}\_600K reproduces equilibrium volumes and thermal expansion trends (\textit{i.e.,} the slope in Figures~\ref{fig:2-benchmark}(d), (e)) within $\le 10$\% $-$ within $\le 5$\% for highly crystalline samples~\cite{liu_tuning_2024} $-$, which is comparable to typical DFT-level agreement.

\begin{table}[tb!]
    \centering
    \renewcommand{\arraystretch}{1.25}
    \begin{tabular}{l | c c c c c c c c}
    \hline \hline
    Models & p & \multicolumn{5}{c}{V [\AA$^3$/atom]} & $E_A$  & $\sigma_{300\text{K}}$\,($\sigma_{\text{min}}$;\,$\sigma_{\text{max}}$)\\
     & [bar] &  0\,K & 200\,K & 300\,K & 400\,K & 600\,K & [eV] & [mS~cm$^{-1}$] \\
    \hline
    \gls{chgnet} & 1 & 22.8 & 23.7 & 25.7 & 26.9 & 31.0 & - & - \\
    \gls{7net}  & 1 & 22.6 & 23.5 & 24.0 & 24.7 & 26.4 & 0.17$\pm$0.01 & 122\,(107;\,183)\\
    \gls{chgnet}\_600K  & 1 & 23.1 & 23.8 & 24.4 & 25.3 & 26.6 & 0.12$\pm$0.01& 331\,(295;\,485)\\
    \gls{chgnet}\_600K  & $10^4$ & -  & 22.2 & 22.6 & 22.9 & 23.9  & 0.24$\pm$0.01 & 15.7\,(13.6;\,23.5)\\
    AIMD~\cite{wang_lithium_2019} & - & 22.4 & -  & - & - & - & 0.19$\pm$0.03 & 14.0\,(4.5;\,47)\\
    exp~\cite{LYB_LYC_exp} & - & - & 21.6 & 21.9 & - & -  & 0.41 & 0.09 \\
    exp*~\cite{LYB_LYC_exp} & - & - & - & - & - & -  & 0.44 & 0.06 \\
    exp**~\cite{LYB_LYC_exp} & - & - & - & - & - & -  & 0.49 & 0.04 \\
    exp\_2~\cite{liu_tuning_2024} & - & - & 21.6 & 21.8 & 22.1 & - & 0.7 & 0.14 \\
    exp\_3~\cite{asano_solid_2018} & - & - & - & 22.0 & - & - & 0.4 & 0.51 \\
    exp\_3*~\cite{asano_solid_2018} & - & - & - & - & - & - & 0.6 & 0.04 \\
   exp\_4~\cite{van_der_maas_investigation_2023} & - & - & - & 22.0 & - & - & 0.7 & 0.05\\
    \hline \hline
    \end{tabular}
    \caption{For the \gls{lyc} system: Predicted volumes (V, in \AA$^3$/atom) at T = 0, 200, 300, 400, and 600\,K, activation energies ($E_A$, in eV), and room-temperature Li conductivities ($\sigma_{300\text{K}}$, in mS~cm$^{-1}$), obtained with \gls{chgnet}, \gls{chgnet}\_600K, and \gls{7net}, and compared to AIMD~\cite{wang_lithium_2019} and experimental data from Refs.~\citenum{LYB_LYC_exp,liu_tuning_2024,van_der_maas_investigation_2023,asano_solid_2018}. We note that these values were partly extracted from figures, when not stated directly in the references.}
    \label{tab:1-lyc}
\end{table}

\begin{table}[tb!]
    \centering
    \renewcommand{\arraystretch}{1.25}
    \begin{tabular}{l | c c c c c c c c}
    \hline \hline
    Models & p & \multicolumn{5}{c}{V [\AA$^3$/atom]} & $E_A$  & $\sigma_{300\text{K}}$\,($\sigma_{\text{min}}$;\,$\sigma_{\text{max}}$) \\
     & [bar] &  0\,K & 200\,K & 300\,K & 400\,K & 600\,K & [eV] & [mS~cm$^{-1}$] \\
    \hline
    \gls{chgnet} & 1 & 29.3 & 33.2 & 34.5 & 37.5 & 42.0 & - & - \\
    \gls{7net} & 1 & 27.6 & 28.9 & 29.6 & 30.2 & 31.9 & 0.20$\pm$0.01 & 42.1\,(33.1;\,79.6) \\
    \gls{chgnet}\_600K & 1 & 27.5 & 28.5 & 29.1 & 29.9 & 31.7 & 0.25$\pm$0.01 & 19.5\,(15.4;\,35.1)\\
    \gls{chgnet}\_600K & $10^4$ & - & 26.6 & 27.0 & 27.4 & 28.6 & 0.36$\pm$0.02 & 0.58\,(0.37;\,1.66) \\
    AIMD~\cite{wang_lithium_2019} & - & 26.9 & - & - & - & - & 0.28$\pm$0.02 & 2.20\,(0.70;\,7.3) \\
    exp~\cite{LYB_LYC_exp} & - & - & 26.1 & 26.8 & - & - & 0.36 & 0.3 \\
    exp*~\cite{LYB_LYC_exp} & - & - & - & - & - & - & 0.34 & 1.4 \\
    exp**~\cite{LYB_LYC_exp} & - & - & - & - & - & - & 0.21 & 4.3 \\
    exp\_3~\cite{asano_solid_2018} & - & - & - & 26.9 & - & - & 0.37 & 0.72 \\
    exp\_3*~\cite{asano_solid_2018} & - & - & - & - & - & - & 0.31 & 1.7 \\
    exp\_4~\cite{van_der_maas_investigation_2023} & - & - & - & - & - & - & 0.33 & 4.7 \\
    \hline \hline
    \end{tabular}
    \caption{For the \gls{lyb} system: Predicted volumes (V, in \AA$^3$/atom) at T = 0, 200, 300, 400, and 600\,K, activation energies ($E_A$, in eV), and room-temperature Li conductivities ($\sigma_{300\text{K}}$, in mS~cm$^{-1}$), obtained with \gls{chgnet}, \gls{chgnet}\_600K, and \gls{7net}, and compared to AIMD~\cite{wang_lithium_2019} and experimental data from Refs.~\citenum{LYB_LYC_exp,asano_solid_2018,van_der_maas_investigation_2023}. We note that these values were partly extracted from figures, when not stated directly in the references.}
    \label{tab:2-lyb}
\end{table}

The remaining $5-10\%$ volume overestimation relative to experiment is expected and can be attributed to two main factors. First, simulated structures correspond to ideal, fully crystalline cells, whereas experimental samples often exhibit lower crystallinity, defects, or microstrain that reduce the measured average volume. Second, the underlying PBE(+U) functional used to generate the training data is known to systematically overestimate equilibrium volumes~\cite{lejaeghere_reproducibility_2016}. Consistent with this interpretation, applying a moderate external pressure of 10\,kbar during \gls{md} simulations brings the predicted volumes into closer agreement with experiment. In subsequent transport analyses, we therefore consider Li-ion dynamics in the $NVT$ ensemble starting from two equilibrated configurations: (1) cells relaxed at 1\,bar $-$ representing fully predictive simulations at the DFT-consistent volume $-$ and (2) cells relaxed at 10\,kbar $-$ yielding volumes closer to experiment and thus more realistic ionic conductivities.

\subsection{Lithium-ion diffusion in \gls{lyc} and \gls{lyb} compounds}

\begin{figure}[tb!]
\centering
\begin{overpic}[width=0.85\linewidth]{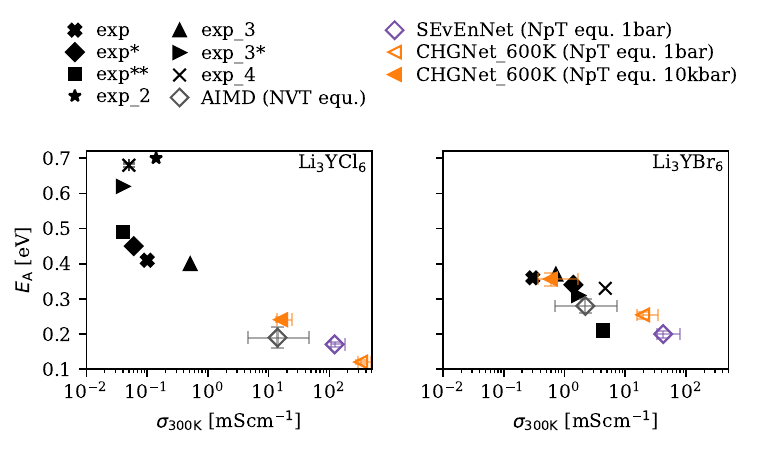}
        \put(10,41){a)} 
        \put(57,41){b)} 
\end{overpic}
\caption{For (a) \gls{lyc} and (b) \gls{lyb}, we report the activation energy ($E_A$) and room-temperature ionic conductivity ($\sigma_{300\text{K}}$), obtained from \gls{md} simulations in the $NVT$ ensemble, using \gls{chgnet}, \gls{chgnet}\_600\,K, and \gls{7net}. The simulations are conducted in $NpT$-equilibrated cells under p=1\,bar (empty symbols) and p=10\,kbar (filled orange triangles). Error bars originate from the variances in the diffusion coefficients that are estimated according to Ref.~\citenum{he_statistical_2018}. We compare our predictions to AIMD~\cite{wang_lithium_2019} and experimental values: exp, exp*, exp** (*different synthesis)~\cite{LYB_LYC_exp}; exp\_2~\cite{liu_tuning_2024}; exp\_3, exp\_3* (*different synthesis)~\cite{asano_solid_2018}; exp\_4~\cite{van_der_maas_investigation_2023}.}
\label{fig:3-conductivity-map}
\end{figure}

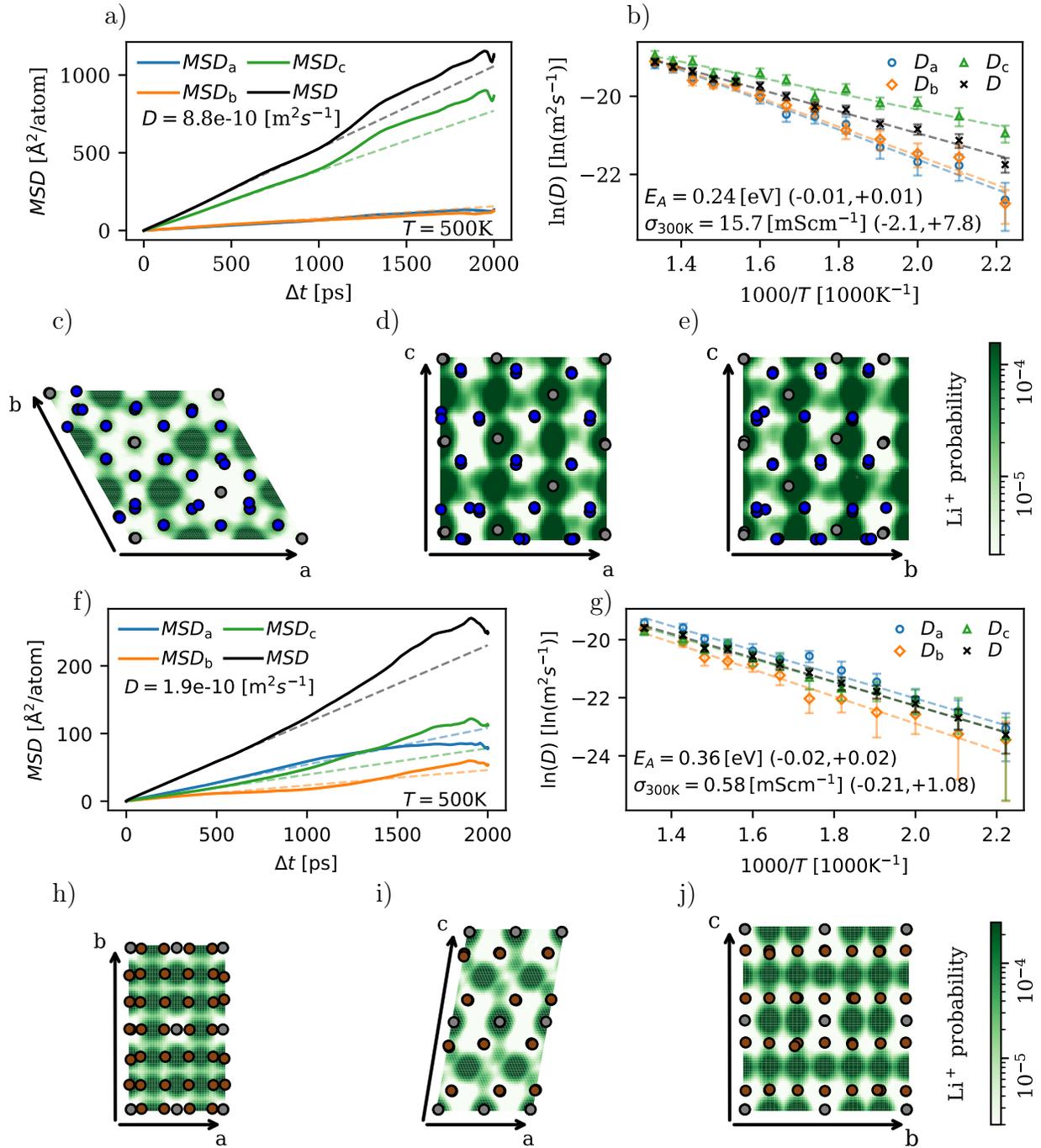
\begin{figure}[ht!]
\centering
\begin{overpic}[width=1.0\linewidth]{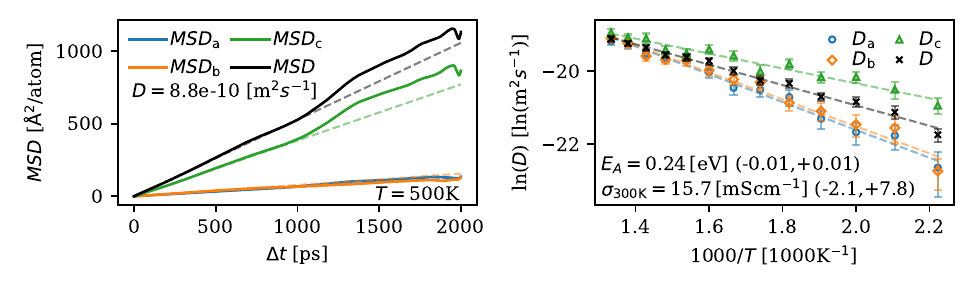}
        \put(10,30){a)} 
        \put(60,30){b)} 
\end{overpic}
\begin{overpic}[width=1.0\linewidth]{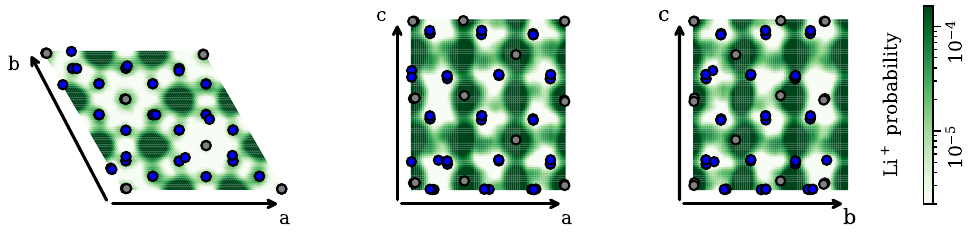}
        \put(5,25){c)} 
        \put(36,25){d)} 
        \put(65,25){e)} 
\end{overpic}
\begin{overpic}[width=1.0\linewidth]{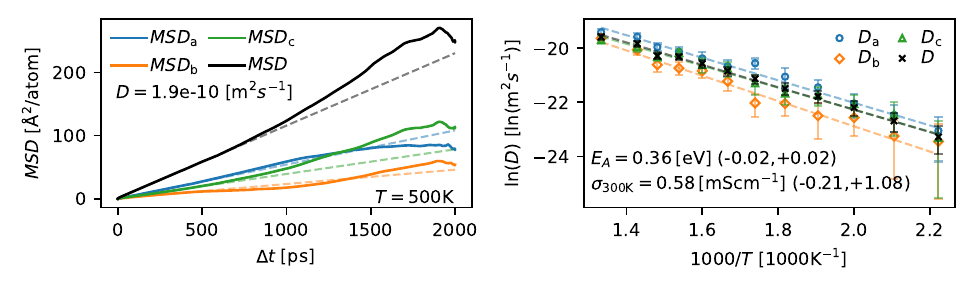}
        \put(7,28.25){f)} 
        \put(56.5,28.25){g)} 
\end{overpic}
\begin{overpic}[width=1.0\linewidth]{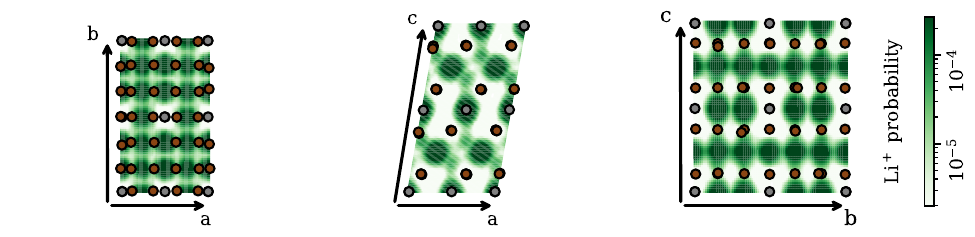}
        \put(5,24.5){h)} 
        \put(36,24.5){i)} 
        \put(65,24.5){j)} 
\end{overpic}
\caption{Li$^+$ transport properties of (a)-(e) \gls{lyc} and (f)-(j) \gls{lyb}, obtained from \gls{md} simulations in the $NVT$ ensemble, starting from $NpT$-equilibrated structures (p=10\,kbar), and using the \gls{chgnet}\_600K potential. (a) and (f) MSD as a function of $\Delta t$ at T=500\,K; (b) and (g) temperature-dependence of the diffusion coefficient, in logarithmic scale; and (c)-(e) and (h)-(j) Li$^+$ probability density maps from \gls{md} simulations at T=500\,K projected on the $ab$-, $ac$-, and $bc$-plane, respectively (Li: green, Y: gray, Cl: blue, Br: brown). Error bars are derived according to the procedure in Ref.~\citenum{he_statistical_2018}.}
\label{fig:4-diffusion}
\end{figure}

Using \gls{chgnet}\_600K, we perform 2\,ns-long $NVT$ \gls{md} simulations in $NpT$-equilibrated cells at temperatures ranging from 400 to 750\,K. From these simulations, we quantify Li-ion transport in \gls{lyc} and \gls{lyb} by extracting self-diffusion coefficients ($D$), activation energies ($E_A$), and room-temperature ionic conductivities ($\sigma_{300\text{K}}$) (Figures~\ref{fig:3-conductivity-map}, \ref{fig:4-diffusion}, and Tables~\ref{tab:1-lyc}, \ref{tab:2-lyb}).

For 1\,bar-equilibrated cells, we obtain $E_A=0.12 \pm 0.01$\,eV ($0.25 \pm 0.01$\,eV) and $\sigma_{300\text{K}}\sim 331$\,mS.cm$^{-1}$ ($\sim 20$\,mS.cm$^{-1}$) for \gls{lyc} (\gls{lyb}), indicated by empty orange triangles in Figure~\ref{fig:3-conductivity-map}. These values are comparable in magnitude to the \gls{7net} predictions and about one order of magnitude higher than the \gls{aimd} results reported in Ref.~\citenum{wang_lithium_2019}, which were obtained from $NVT$ simulations using 0\,K DFT optimized cells (without $NpT$ cell equilibration) $-$ effectively imposing a large internal pressure. When using cells compressed to near-experimental volumes (p=10\,kbar, filled orange triangles), Li-ion mobility decreases, yielding $E_A = 0.24 \pm 0.01$\,eV ($0.36 \pm 0.02$\,eV) and $\sigma_{300\text{K}} \sim 15.7$\,mS.cm$^{-1}$ ($\sim 0.6$\,mS.cm$^{-1}$) for \gls{lyc} (\gls{lyb}), in much better agreement with the \gls{aimd} data. Simulated conductivities for \gls{lyc} exceed experimental values obtained via electrochemical impedance spectroscopy near 300\,K~\cite{asano_solid_2018,LYB_LYC_exp,van_der_maas_investigation_2023,liu_tuning_2024}, reflecting an overestimation when extrapolating higher-temperature diffusivities to room temperature. This discrepancy is consistent with the two diffusion regimes observed experimentally~\cite{liu_tuning_2024}: a high-T regime ($T>\qty{405}{K}$, $E\tief{A}=\qty{0.22}{eV}$) and a low-T regime ($T<\qty{405}{K}$, $E\tief{A}=\qty{0.7}{eV}$). For LYB, by contrast, the \gls{chgnet}\_600K predictions for compressed cells align well with both \gls{aimd} results and experiments.\\

Analysis of \glspl{msd} and diffusivities reveals a clear contrast in Li-ion transport between \gls{lyc} and \gls{lyb} (Figure~\ref{fig:4-diffusion}). In \gls{lyc}, Li$^+$ diffusion is highly anisotropic, with diffusion coefficients at 500\,K approximately $2-3$ times larger along the $c$ axis than within the $ab$-plane (Figures~\ref{fig:4-diffusion}(a), (b)). Li$^+$ probability density maps (Figures~\ref{fig:4-diffusion}(c)-(e)) reveal preferential migration through adjacent face-sharing octahedral sites along $c$ over edge-sharing pathways in the $ab$-plane, with an intermediate tetrahedral site. This behavior is consistent with the layered structure providing open diffusion channels along $c$. In contrast, \gls{lyb} exhibits nearly isotropic diffusion with a diffusion coefficient of $1.9 \times 10^{-10}$\,m$^{2}$~s$^{-1}$ at 500\,K. This isotropy reflects migration occurring exclusively via edge-sharing pathways with an intermediate tetrahedral site (Figures~\ref{fig:4-diffusion}(f)-(j)).

\subsection{Effect of halide substitution on phase stability and Li-ion transport}

Last, we investigate the impact of Cl/Br substitution on phase stability and Li$^+$ transport across the mixed \gls{lycb} series. The phase stability is assessed from the total energy difference between the two competing phases $-$ trigonal $P\bar{3}c1$ and monoclinic $C2$. Table~\ref{tab2} lists the calculated energy difference, $\Delta E = E_{P\bar{3}c1}-E_{C2}$, for $\text{x}=0$ to 6, obtained from \textit{ab initio} calculations (for details, see Figure~S8). Comparable trends are reproduced using \gls{chgnet}\_600K. The results reveal that the two phases are nearly degenerate, with energy differences of only a few meV/atom, suggesting the possibility of solid-solution behavior at finite temperatures due to thermal contributions. At Br-rich compositions ($\text{x}>1.5$), the monoclinic $C2$ phase becomes thermodynamically favored, in agreement with the experimentally observed phase transition in the range $1.5<\text{x}\le3$ reported in Ref.~\citenum{van_der_maas_investigation_2023}.

\begin{table}[ht!]
    \centering
    \begin{tabular}{c c@{\hspace*{1.5cm}} c c}
    \hline \hline
    x$_{\text{Br}}$ & $\Delta E$ & x$_{\text{Br}}$ & $\Delta E$ \\
     & (meV/atom) & & (meV/atom) \\
    \hline
    0.0 & $-2.27$ & 3.5 & $\phantom{+}4.30$ \\
    0.5 & $-2.08$ & 4.0 & $\phantom{+}4.51$ \\
    1.0 & $-0.82$ & 4.5 & $\phantom{+}4.41$ \\
    1.5 & $-0.25$ & 5.0 & $\phantom{+}4.15$ \\
    2.0 & $\phantom{+}1.79$ & 5.5 & $\phantom{+}4.62$ \\
    2.5 & $\phantom{+}3.28$ & 6.0 & $\phantom{+}5.22$ \\
    3.0 & $\phantom{+}3.88$ &  & \\
    \hline \hline
    \end{tabular}
    \caption{Differences in total energy between the two phases, $\Delta E = E_{P\bar{3}c1}-E_{C2}$ for $\text{x}=0$ to 6, expressed in meV/atom, obtained from DFT structure optimization. Negative $\Delta E$ indicate the $P\bar{3}c1$ phase is energetically favored, whereas positive $\Delta E$ mean that $C2$ is more stable. We note that these values may slightly vary depending on the ML model used in the structure enumeration and ranking procedure.}
    \label{tab2}
\end{table}

\begin{figure}[tb!]
\centering
\begin{overpic}[width=1.0\linewidth]{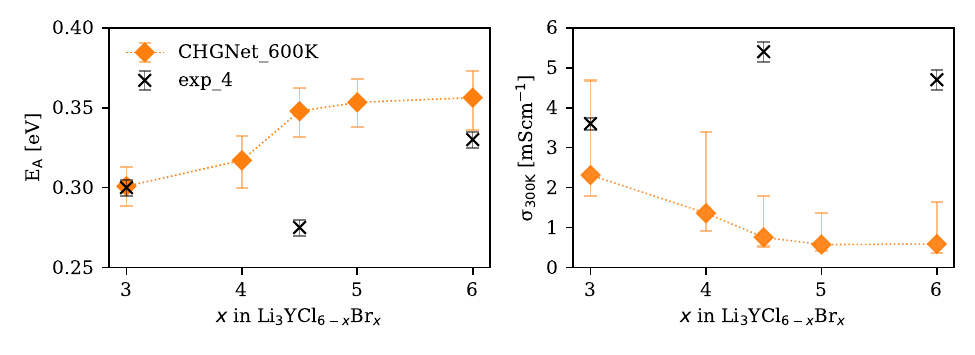}
        \put(11,35.5){a)} 
        \put(59,35.5){b)} 
\end{overpic}
\caption{Li$^+$ transport properties for \gls{lycb} obtained from \gls{chgnet}\_600K \gls{md} simulations (starting from $NpT$-equilibrated structures under p=10\,kbar), including (a) activation energy $E_A$ and (b) room-temperature ionic conductivity $\sigma_{300\text{K}}$. We consider $C2$-symmetry structures of \gls{lycb} with $\text{x}=3$ to 6. Experimental values (exp\_4~\cite{van_der_maas_investigation_2023}) are shown with black crosses.}
\label{fig:5-conductivity-lycb}
\end{figure}

Figure~\ref{fig:5-conductivity-lycb} presents Li$^+$ transport properties for various Br-rich \gls{lycb} compositions in the $C2$ crystal structure. As the Cl content increases (\textit{i.e.}, $\text{x}=6$ down to 3), the room-temperature Li$^+$ conductivity initially remains roughly constant around 0.7\,mS.cm$^{-1}$, before increasing monotonically for $5.0\ge\text{x}\ge3.0$. This trend is inversely reflected in the activation energies and can be attributed to increased disorder and strain in the lattice, which lowers diffusion barriers. Although the experimentally reported conductivity maximum between $\text{x}=3$ and $\text{x}=6$ is not fully reproduced~\cite{van_der_maas_investigation_2023}, the predicted values are close to experiment and indicate that $\sigma_{300\text{K}}$ of LYB can be enhanced through Br$\,\rightarrow\,$Cl substitution. Analysis of \glspl{msd}, diffusion coefficients, and Li$^+$ probability density maps for Li$_{3}$YCl$_{3}$Br$_{3}$ in Figure~S9 shows isotropic transport behavior, qualitatively similar to the parent \gls{lyb} system (with the same $C2$ parent space group).

\section{Conclusions}
\label{sec:Conclusion}

In summary, this work establishes a data-efficient and transferable framework for modeling lithium transport in halide \glspl{se} via targeted fine-tuning of \glspl{umlip}. Starting from experimentally refined disordered structures, we systematically enumerate and rank ordered configurations to identify physically meaningful low-energy models, which are then equilibrated under realistic thermodynamic conditions using MLIP-driven simulations. Fine-tuning the \gls{chgnet} model on halide-specific data restores near-DFT accuracy in energy, force, and stress tensor predictions, stabilize $NpT$ simulations up to 800\,K, and enables nanosecond-scale \gls{md} simulations of Li$^+$ diffusion across the full \gls{lycb} composition range.

The fine-tuned \gls{chgnet} model achieves quantitative agreement with \gls{dft} and \gls{aimd} at a cost reduction exceeding four orders of magnitude, and remains transferable across the entire \gls{lycb} series. Beyond reproducing known behavior, it provides key insights into anisotropic Li$^+$ migration in \gls{lyc}, isotropic transport in \gls{lyb}, and the pressure- and composition-dependent modulation of ionic conductivity. More broadly, this work demonstrates that targeted fine-tuning of \glspl{umlip} bridges the gap between efficiency and fidelity, providing a practical route for predictive modeling of complex, partially disordered materials that remain inaccessible to high-throughput \gls{dft} databases. Overall, the presented approach lays a groundwork for accelerating the discovery and optimization of next-generation \glspl{se} through integrated exploration of structure, properties, and transport.


\section{Computational Methods}
\label{sec:Experimental Section}

\textbf{\gls{dft} calculations $-$} \gls{dft} calculations are conducted with the \gls{vasp} ~\cite{VASP,VASP2} using the \gls{gga} as proposed by \gls{pbe}~\cite{GGA_PBE}. The static calculations used to produce the training dataset are performed using a $3\times3\times3$ $\Gamma$-centered $\mathbf{k}$-point grid and a cutoff energy of 350\,eV, ensuring energy is converged within \qty{0.5}{meV.atom}$^{-1}$. For structure optimization, we use a $\Gamma$-centered $\mathbf{k}$-point grid, with a $\mathbf{k}$-point-spacing of \qty{0.2}{\angstrom}$^{-1}$, and a cutoff energy of 520\,eV. Atomic positions and cell parameters are optimized with the FIRE~\cite{fire} optimizer and a convergence threshold of \qty{0.01}{eV\angstrom}$^{-1}$. The same optimizer and threshold are used for \gls{mlip}-based structure optimization. For the AIMD simulations, we use a timestep of \qty{1}{fs}, a cutoff energy of 350\,eV, a Nosé-Hoover thermostat in the $NVT$ ensemble, and a $2\times2\times2$ $\Gamma$-centered $\mathbf{k}$-point grid. Finite-temperature volumes are predicted using an equation of state approach, such that the initial cell volume is scaled by 0\%, 5\%, and 10\%, and the resulting volumes are fitted to the corresponding $NVT$-equilibrated pressures (averaged over at least \qty{1}{ps} after equilibration) using a quadratic function.

\textbf{Structure enumeration and ranking $-$} We start from the experimental LYC and LYB structures~\cite{LYB_LYC_exp} and build $1 \times 1 \times 2$ supercells containing 60 and 40 atoms, respectively. We enumerate all symmetrically distinct configurations using \textit{enumlib}~\cite{enum}. Total energy and formation energy are predicted with the \textit{matgl} package~\cite{matgl}, using algorithms implemented in ASE~\cite{ASE} and different \glspl{umlip} as calculators. \gls{m3gnet} and \gls{chgnet} provide similar energy predictions, with the former being about 3 times faster $-$ hence \gls{m3gnet} is used for the ranking procedure here. Structure optimization is performed using the ASE optimizer, employing the FIRE algorithm and a convergence threshold of \qty{0.01}{eV\angstrom}$^{-1}$. The results reported herein are generated using the following software versions: Python 3.10.10, Pymatgen 2025.2.18, PyTorch 2.4.0+cu124, ASE 3.23.0, \gls{m3gnet} 0.2.4, \gls{chgnet} 0.3.8 (pretrained model \gls{chgnet} 0.3.0), sevenn 0.11.0 (pretrained model SEvEnNet-MF-ompa~\cite{7net_mf_ompa}).

\textbf{\gls{umlip} fine-tuning procedure $-$} The pretrained \gls{chgnet} \gls{umlip} is fine-tuned to enable stable $NpT$ \gls{md} simulations of \gls{lycb} structures at temperatures up to 800\,K. The following iterative procedure is used (Figure~\ref{fig:1-workflow}):
\setlist{nolistsep}
\begin{enumerate}[label=\textbf{\arabic*)}, leftmargin=*, align=left,noitemsep]
  \item Generate finite-temperature structures at $T_i$:
  \begin{enumerate}[label=\textbf{1.\arabic*)}, leftmargin=*, align=left]
  \item Starting from the LYC0 and LYB0 structures, equilibrate the cell volume by $NpT$ \gls{md} at $T_i$ and 1\,bar, using \gls{chgnet}\_$T_{i-1}$). Subsequently perform an $NVT$ \gls{md} production run (1,000,000 steps; store configurations every 10th step).
  \item For each system, derive a subset of representative structures from the trajectory to ensure broad geometric diversity. Selection is performed using furthest-point sampling based on the mean-squared error in atomic positions.
  \end{enumerate}
  \item Calculate energy, force, and stress labels using static \gls{dft}.
  \item Fine-tune the \gls{chgnet} model with the labeled dataset, composed of 50\% \gls{lyc} and 50\% \gls{lyb} structures, yielding the fine-tuned \gls{chgnet}\_$T_i$ model. The batch size, number of epochs, initial learning rate, and number of frozen layers were optimized for this step, while all other hyperparameters were kept as in Ref.~\citenum{chgnet}.
  \item If $T_i< T_{max}$, the temperature is increased according to $T_{i+1}= T_i+\Delta T$, and the procedure is repeated from step \textbf{1.2)}.
\end{enumerate}
This iterative strategy ensures that the structural configurations encountered during \gls{md} remain close to those represented in the progressively expanding training dataset, thereby maintaining stability and physical reliability of the simulations. The temperature increment $\Delta T$ controls the magnitude of structural perturbations between iterations and was set to 200\,K. Step \textbf{1)} required approximately $0.8$\,s/atom/picosecond on a GPU, while step \textbf{3)} required $\sim 5-10$\,minutes on a GPU for a dataset containing $150-300$ structures per iteration. In each iteration, the model was fine-tuned incrementally using only the dataset generated at the current temperature, starting from the model obtained in the previous step.

\textbf{\gls{umlip}-driven \gls{md} simulations $-$} \gls{md} simulations are performed with the GPU-accelerated LAMMPS package~\cite{lammps, lammps_gpu_1,lammps_gpu_2,lammps_gpu_3,lammps_gpu_4,lammps_gpu_5,lammps_gpu_6}, using either the (finetuned) \gls{chgnet}~\cite{chgnet,lammps_chgnet_m3gnet} or \gls{7net}-MF-ompa~\cite{7net} \gls{umlip}. A timestep of 1\,fs is used. Temperature coupling for $NVT$ and $NpT$ simulations is set to 0.1\,ps andpressure coupling (for $NpT$) to 1\,ps using the Nosé-Hoover thermostat and barostat~\cite{nose_molecular_1984,hoover_canonical_1985}. Prior to each $NVT$ production run, the following equilibration procedure is carried out: velocities are initialized at 100\,K and the system is linearly heated to the target temperature over 2\,ps while keeping the cell parameters fixed. The system is then equilibrated for 50\,ps in the $NVT$ ensemble. For the $NpT$-equilibrated cells, this step is followed by an additional 140\,ps equilibration at a target pressure of 1\,bar. The final cell is defined by the average lattice parameters and volume over the last 112\,ps of the $NpT$ equilibration. For the 60- and 40-atoms LYC and LYB supercells, the computational cost on a GPU is approximately 0.8\,s/atom/picosecond, \textit{i.e.}, more than $10^5$ times faster than a comparable \gls{aimd} simulation performed on a CPU. 

\textbf{Ionic conductivity derivation $-$} From an \gls{md} trajectory of duration $t$ at temperature $T$, ion transport properties including the self-diffusion coefficient, $D$, ionic conductivity, $\sigma$, and activation energy for diffusion, $E_A$, are obtained. First, the average atomic \gls{msd} over a time interval $\Delta t$ for $N$ atoms $i$ is computed as~\cite{ong_phase_2013}:
\begin{equation}
  \label{eq:MSD_ave}
\text{MSD}(\Delta t)=\left\langle |\mathbf{r}(\Delta t)|^2 \right\rangle = \frac{1}{N} \sum\lo{i} \left\langle \left| \mathbf{r}\lo{i}(t\lo{0} + \Delta t) - \mathbf{r}\lo{i}(t\lo{0}) \right|^2 \right\rangle\lo{t\lo{0}}
\end{equation}
where $\Delta t$ is a time interval with an upper bound given by the total simulation time $t$, $t\lo{0}$ is a given timestep with $t_0 \in [0,t-\Delta t]$, and $\left\langle\right\rangle\lo{t\lo{0}}$ indicates an average over all possible $t\lo{0}$. The lithium self-diffusion coefficient $D$ at temperature $T$ is obtained from the Einstein relation~\cite{baktash_diffusion_2020}:
\begin{equation}
  \label{eq:einstein_D}
  D = \lim\lo{\Delta t \to \infty} \frac{MSD(\Delta t)}{2d\Delta t}
\end{equation}
where $d$ is the dimensionality of diffusion (\textit{e.g.}, $d=3$ for 3D diffusion). In practice, $D$ is determined from a linear fit to discrete \gls{msd}$(\Delta t)$ data. To ensure a reliable fit, the ballistic diffusion regime at short times ($0<\Delta t<x\,ps$) and the region with poor statistics at large lag times ($\Delta t>0.7t$) are excluded, following Ref~\citenum{he_statistical_2018}. As the calculated \gls{msd}$(\Delta t)$ exhibits deviations from linear behavior beyond approximately $500-1000$\,ps, the maximum fitting window was limited to $\Delta t_{max}=500$\,ps. Error bars in $D$ are estimated according to the procedure described in Ref.~\citenum{he_statistical_2018}. Activation energies for diffusion are obtained by fitting $\ln{(D)}$ versus $1/T$ over multiple temperature using the Arrhenius relation:
\begin{equation}
  \label{eq:EA_diff}
  D(T) = \tilde{D}\exp\left({-\dfrac{E_A}{k_B}T}\right )
\end{equation}
where $\tilde{D}$ is the constant diffusion coefficient for $T\rightarrow\infty$. If $N$ mobile ions of charge $q$ are present in a system of volume $V$, the ionic conductivity $\sigma$ is calculated via the Nernst-Einstein relation~\cite{baktash_diffusion_2020,evans_statistical_2007}:
\begin{equation}
  \label{eq:einstein_sig}
  \sigma(T) = \dfrac{Nq^2}{Vk_BT}\cdot D(T).
\end{equation}

\begin{suppinfo}
The Supplementary Information is available free of charge.
\begin{itemize}
    \item Fine-tuning dataset at finie-temperature; Transferability across various test sets; Alternative accumulative fine-tuning approach; Optimization of cell parameters; Formation energies and phase stability prediction; Lithium-ion diffusion in Li$_3$YCl$_3$Br$_3$ (PDF).
    \item Training dataset; all fine-tuned CHGNet models; low-energy LYCB structures from ranking (zip).
\end{itemize}
\end{suppinfo}

\begin{acknowledgement}
The authors thank Francesco Ricci, Yusuf Shaidu, and Dany Carlier for stimulating discussions, Zhenming Xu for providing additional information on Ref~\citenum{xu_machine_2023}, Bassem Sboui for computational support, and Miguel Ceja for support with the Vesta package. The calculations were primarily supported by the computing facilities provided by the Mésocentre de Calcul Intensif Aquitain (MCIA) of the University of Bordeaux. The authors thank the French National Research Agency (STORE-EX Labex Project ANR-10-LABX-76-01) for financial support.
\end{acknowledgement}

\bibliography{biblio.bib}

\end{document}

%% file: preamble.tex
\usepackage{comment}
\usepackage[acronyms]{glossaries}%
\usepackage{amsmath}

\newcommand{\lo}[1]{_{\mathrm{#1}}}
\usepackage{siunitx}
\usepackage{float}
\usepackage{enumitem} 

\newcommand{\tief}[1]{_{\mathrm{#1}}}

\newcommand{\arr}[1]{[#1]}
\usepackage[normalem]{ulem}
\usepackage[percent]{overpic} 

\usepackage{booktabs}
\usepackage{chemformula} 
\usepackage[T1]{fontenc} 
\usepackage{braket}
\usepackage[colorlinks=true,linkcolor=blue,citecolor=blue,urlcolor=blue]{hyperref}

%% file: abbreviations.tex
\makenoidxglossaries%
\newacronym{md}{MD}{molecular dynamics}%
\newacronym{aimd}{AIMD}{ab-initio molecular dynamics}%
\newacronym{mc}{MC}{Monte Carlo}%
\newacronym{msd}{MSD}{mean-squared displacement}%
\newacronym{rmse}{MSD}{root mean square error}%
\newacronym{lycb}{LYCB}{Li$\lo{3}$YCl$\lo{6-x}$Br$\lo{x}$}%
\newacronym{lyb}{LYB}{Li$\lo{3}$YBr$\lo{6}$}%
\newacronym{lyc}{LYC}{Li$\lo{3}$YCl$\lo{6}$}%
\newacronym{scf}{SCF}{self consistent field}%
\newacronym{ml}{ML}{machine learning}%
\newacronym{vasp}{VASP}{Vienna Ab initio Simulation Package}%
\newacronym{mae}{MAE}{mean average error}%
\newacronym{pbe}{PBE}{Perdew–Burke–Ernzerhof}%
\newacronym{gmlp}{gMLP}{gated multy layer perceptron}%
\newacronym{nn}{NN}{neural network}%
\newacronym{gnn}{GNN}{graph neural network}%
\newacronym{gcn}{GCN}{graph convolution network}%
\newacronym{pes}{PES}{potential energy surface}%
\newacronym{ks}{KS}{Kohn-Sham}%
\newacronym{mlip}{MLIP}{Machine learning interatomic potential}%
\newacronym{umlip}{uMLIP}{universal machine learning interatomic potential}%
\newacronym{dft}{DFT}{density functional theory}%
\newacronym{lda}{LDA}{local density approximation}%
\newacronym{gga}{GGA}{generalized gradient approximation}%
\newacronym{assb}{ASSB}{all solid state battery}%
\newacronym{assbs}{ASSBs}{all solid state batteries}%
\newacronym{se}{SE}{solid electrolyte}%
\newacronym{m3gnet}{M3GNet}{Materials 3-body Graph Network}%
\newacronym{chgnet}{CHGNet}{Crystal Hamiltonian Graph Network}%
\newacronym{ch800}{CHGNet\_800}{CHGNet iteratively finetuned on MD data up to 800\,K}%
\newacronym{7net}{SevenNet}{Scalable EquiVariance-Enabled Neural Network}%
\newacronym{nequip}{NequIP}{Neural equivariant Interatomic Potential}%
\newacronym{fire}{FIRE}{Fast Inertial Relaxation Engine}%
\newacronym{xrd}{XRD}{X-Ray diffraction}%

%% file: biblio.bib
@article{ewald_berechnung_1921,
	title = {Die {Berechnung} optischer und elektrostatischer {Gitterpotentiale}},
	volume = {369},
	issn = {0003-3804, 1521-3889},
	url = {https://onlinelibrary.wiley.com/doi/10.1002/andp.19213690304},
	doi = {10.1002/andp.19213690304},
	language = {en},
	number = {3},
	urldate = {2025-10-31},
	journal = {Annalen der Physik},
	author = {Ewald, P. P.},
	month = jan,
	year = {1921},
	pages = {253--287},
}

@article{toukmaji_ewald_1996,
	title = {Ewald summation techniques in perspective: a survey},
	volume = {95},
	copyright = {https://www.elsevier.com/tdm/userlicense/1.0/},
	issn = {00104655},
	shorttitle = {Ewald summation techniques in perspective},
	url = {https://linkinghub.elsevier.com/retrieve/pii/0010465596000161},
	doi = {10.1016/0010-4655(96)00016-1},
	language = {en},
	number = {2-3},
	urldate = {2025-10-31},
	journal = {Computer Physics Communications},
	author = {Toukmaji, Abdulnour Y. and Board, John A.},
	month = jun,
	year = {1996},
	pages = {73--92},
}

@article{radova_fine-tuning_2025,
	title = {Fine-tuning foundation models of materials interatomic potentials with frozen transfer learning},
	volume = {11},
	issn = {2057-3960},
	url = {https://www.nature.com/articles/s41524-025-01727-x},
	doi = {10.1038/s41524-025-01727-x},
	language = {en},
	number = {1},
	urldate = {2025-08-17},
	journal = {npj Computational Materials},
	author = {Radova, Mariia and Stark, Wojciech G. and Allen, Connor S. and Maurer, Reinhard J. and Bartók, Albert P.},
	month = jul,
	year = {2025},
	pages = {237},
}

@article{kaur_data-efficient_2025,
	title = {Data-efficient fine-tuning of foundational models for first-principles quality sublimation enthalpies},
	volume = {256},
	issn = {1359-6640, 1364-5498},
	url = {https://xlink.rsc.org/?DOI=D4FD00107A},
	doi = {10.1039/D4FD00107A},
	language = {en},
	urldate = {2025-08-17},
	journal = {Faraday Discussions},
	author = {Kaur, Harveen and Della Pia, Flaviano and Batatia, Ilyes and Advincula, Xavier R. and Shi, Benjamin X. and Lan, Jinggang and Csányi, Gábor and Michaelides, Angelos and Kapil, Venkat},
	year = {2025},
	pages = {120--138},
}

@misc{liu_study_2025,
	title = {A {Study} on the {Fine}-{Tuning} {Performance} of {Universal} {Machine}-{Learned} {Interatomic} {Potentials} ({U}-{MLIPs})},
	copyright = {Creative Commons Zero v1.0 Universal},
	url = {https://arxiv.org/abs/2506.07401},
	doi = {10.48550/ARXIV.2506.07401},
	urldate = {2025-08-17},
	publisher = {arXiv},
	author = {Liu, Xiaoqing and Zeng, Kehan and Wang, Yangshuai and Zhao, Teng},
	year = {2025},
	note = {Version Number: 1},
}

@misc{huang_cross-functional_2025,
	title = {Cross-functional transferability in universal machine learning interatomic potentials},
	copyright = {Creative Commons Attribution 4.0 International},
	url = {https://arxiv.org/abs/2504.05565},
	doi = {10.48550/ARXIV.2504.05565},
	urldate = {2025-08-17},
	publisher = {arXiv},
	author = {Huang, Xu and Deng, Bowen and Zhong, Peichen and Kaplan, Aaron D. and Persson, Kristin A. and Ceder, Gerbrand},
	year = {2025},
	note = {Version Number: 1},
}

@article{hoover_canonical_1985,
	title = {Canonical dynamics: {Equilibrium} phase-space distributions},
	volume = {31},
	copyright = {http://link.aps.org/licenses/aps-default-license},
	issn = {0556-2791},
	shorttitle = {Canonical dynamics},
	url = {https://link.aps.org/doi/10.1103/PhysRevA.31.1695},
	doi = {10.1103/PhysRevA.31.1695},
	language = {en},
	number = {3},
	urldate = {2024-05-12},
	journal = {Physical Review A},
	author = {Hoover, William G.},
	month = mar,
	year = {1985},
	pages = {1695--1697},
}

@article{nose_molecular_1984,
	title = {A molecular dynamics method for simulations in the canonical ensemble},
	volume = {52},
	issn = {0026-8976, 1362-3028},
	url = {http://www.tandfonline.com/doi/abs/10.1080/00268978400101201},
	doi = {10.1080/00268978400101201},
	pages = {255--268},
	number = {2},
	journal = {Molecular Physics},
	shortjournal = {Molecular Physics},
	author = {Nosé, Shūichi},
	urlyear = {2024-02-14},
	year = {1984},
	langid = {english},
}

@article{wang_machine_2024,
	title = {Machine learning interatomic potential: {Bridge} the gap between small-scale models and realistic device-scale simulations},
	volume = {27},
	issn = {25890042},
	shorttitle = {Machine learning interatomic potential},
	url = {https://linkinghub.elsevier.com/retrieve/pii/S2589004224008952},
	doi = {10.1016/j.isci.2024.109673},
	language = {en},
	number = {5},
	urldate = {2025-05-06},
	journal = {iScience},
	author = {Wang, Guanjie and Wang, Changrui and Zhang, Xuanguang and Li, Zefeng and Zhou, Jian and Sun, Zhimei},
	month = may,
	year = {2024},
	pages = {109673},
}

@article{jacobs_practical_2025,
	title = {A practical guide to machine learning interatomic potentials – {Status} and future},
	volume = {35},
	issn = {13590286},
	url = {https://linkinghub.elsevier.com/retrieve/pii/S1359028625000014},
	doi = {10.1016/j.cossms.2025.101214},
	language = {en},
	urldate = {2025-05-06},
	journal = {Current Opinion in Solid State and Materials Science},
	author = {Jacobs, Ryan and Morgan, Dane and Attarian, Siamak and Meng, Jun and Shen, Chen and Wu, Zhenghao and Xie, Clare Yijia and Yang, Julia H. and Artrith, Nongnuch and Blaiszik, Ben and Ceder, Gerbrand and Choudhary, Kamal and Csanyi, Gabor and Cubuk, Ekin Dogus and Deng, Bowen and Drautz, Ralf and Fu, Xiang and Godwin, Jonathan and Honavar, Vasant and Isayev, Olexandr and Johansson, Anders and Kozinsky, Boris and Martiniani, Stefano and Ong, Shyue Ping and Poltavsky, Igor and Schmidt, Kj and Takamoto, So and Thompson, Aidan P. and Westermayr, Julia and Wood, Brandon M.},
	month = mar,
	year = {2025},
	pages = {101214},
}

@article{m3gnet,
	title = {A universal graph deep learning interatomic potential for the periodic table},
	volume = {2},
	issn = {2662-8457},
	url = {https://www.nature.com/articles/s43588-022-00349-3},
	doi = {10.1038/s43588-022-00349-3},
	language = {en},
	number = {11},
	urldate = {2025-05-06},
	journal = {Nature Computational Science},
	author = {Chen, Chi and Ong, Shyue Ping},
	month = nov,
	year = {2022},
	pages = {718--728},
}

@article{chgnet,
	title = {{CHGNet} as a pretrained universal neural network potential for charge-informed atomistic modelling},
	volume = {5},
	issn = {2522-5839},
	url = {https://www.nature.com/articles/s42256-023-00716-3},
	doi = {10.1038/s42256-023-00716-3},
	language = {en},
	number = {9},
	urldate = {2025-05-06},
	journal = {Nature Machine Intelligence},
	author = {Deng, Bowen and Zhong, Peichen and Jun, KyuJung and Riebesell, Janosh and Han, Kevin and Bartel, Christopher J. and Ceder, Gerbrand},
	month = sep,
	year = {2023},
	pages = {1031--1041},
}

@article{mat_proj,
	title = {Commentary: {The} {Materials} {Project}: {A} materials genome approach to accelerating materials innovation},
	volume = {1},
	issn = {2166-532X},
	shorttitle = {Commentary},
	url = {https://pubs.aip.org/apm/article/1/1/011002/119685/Commentary-The-Materials-Project-A-materials},
	doi = {10.1063/1.4812323},
	language = {en},
	number = {1},
	urldate = {2025-05-08},
	journal = {APL Materials},
	author = {Jain, Anubhav and Ong, Shyue Ping and Hautier, Geoffroy and Chen, Wei and Richards, William Davidson and Dacek, Stephen and Cholia, Shreyas and Gunter, Dan and Skinner, David and Ceder, Gerbrand and Persson, Kristin A.},
	month = jul,
	year = {2013},
	pages = {011002},
}

@misc{barroso_omat24,
      title={Open Materials 2024 (OMat24) Inorganic Materials Dataset and Models}, 
      author={Luis Barroso-Luque and Muhammed Shuaibi and Xiang Fu and Brandon M. Wood and Misko Dzamba and Meng Gao and Ammar Rizvi and C. Lawrence Zitnick and Zachary W. Ulissi},
      year={2024},
      eprint={2410.12771},
      archivePrefix={arXiv},
      primaryClass={cond-mat.mtrl-sci},
      url={https://arxiv.org/abs/2410.12771}, 
}

@misc{mat_proj_traj,
	title = {Materials {Project} {Trajectory} ({MPtrj}) {Dataset}},
	copyright = {MIT License},
	url = {https://figshare.com/articles/dataset/Materials_Project_Trjectory_MPtrj_Dataset/23713842/2},
	doi = {10.6084/M9.FIGSHARE.23713842.V2},
	urldate = {2025-05-08},
	publisher = {figshare},
	author = {Deng, Bowen},
	year = {2023},
	note = {Artwork Size: 12188168685 Bytes
Pages: 12188168685 Bytes},
}

@article{nequip,
	title = {E(3)-equivariant graph neural networks for data-efficient and accurate interatomic potentials},
	volume = {13},
	issn = {2041-1723},
	url = {https://www.nature.com/articles/s41467-022-29939-5},
	doi = {10.1038/s41467-022-29939-5},
	language = {en},
	number = {1},
	urldate = {2025-05-08},
	journal = {Nature Communications},
	author = {Batzner, Simon and Musaelian, Albert and Sun, Lixin and Geiger, Mario and Mailoa, Jonathan P. and Kornbluth, Mordechai and Molinari, Nicola and Smidt, Tess E. and Kozinsky, Boris},
	month = may,
	year = {2022},
	pages = {2453},
}

@article{fire,
  title={Structural relaxation made simple},
  author={Bitzek, Erik and Koskinen, Pekka and G{\"a}hler, Franz and Moseler, Michael and Gumbsch, Peter},
  journal={Physical Review Letters},
  volume={97},
  number={17},
  pages={170201},
  year={2006},
  publisher={American Physical Society},
  doi={10.1103/PhysRevLett.97.170201}
}

@misc{matgl,
	title = {Materials {Graph} {Library} ({MatGL}), an open-source graph deep learning library for materials science and chemistry},
	copyright = {Creative Commons Attribution 4.0 International},
	url = {https://arxiv.org/abs/2503.03837},
	doi = {10.48550/ARXIV.2503.03837},
	urldate = {2025-05-16},
	publisher = {arXiv},
	author = {Ko, Tsz Wai and Deng, Bowen and Nassar, Marcel and Barroso-Luque, Luis and Liu, Runze and Qi, Ji and Liu, Elliott and Ceder, Gerbrand and Miret, Santiago and Ong, Shyue Ping},
	year = {2025},
	note = {Version Number: 1},
}

@article{enum,
	title = {Algorithm for generating derivative structures},
	volume = {77},
	copyright = {http://link.aps.org/licenses/aps-default-license},
	issn = {1098-0121, 1550-235X},
	url = {https://link.aps.org/doi/10.1103/PhysRevB.77.224115},
	doi = {10.1103/PhysRevB.77.224115},
	language = {en},
	number = {22},
	urldate = {2025-05-16},
	journal = {Physical Review B},
	author = {Hart, Gus L. W. and Forcade, Rodney W.},
	month = jun,
	year = {2008},
	pages = {224115},
}

@misc{lammps_chgnet_m3gnet,
  author       = {AdvanceSoft Corporation},
  title        = {Customized LAMMPS (2Aug2023) for Neural Network Potential},
  year         = {2023},
  howpublished = {\url{https://github.com/advancesoftcorp/lammps}},
  note         = {Accessed: 2025-05-16}
}

@article{lammps,
	title = {{LAMMPS} - a flexible simulation tool for particle-based materials modeling at the atomic, meso, and continuum scales},
	volume = {271},
	issn = {00104655},
	url = {https://linkinghub.elsevier.com/retrieve/pii/S0010465521002836},
	doi = {10.1016/j.cpc.2021.108171},
	language = {en},
	urldate = {2025-05-16},
	journal = {Computer Physics Communications},
	author = {Thompson, Aidan P. and Aktulga, H. Metin and Berger, Richard and Bolintineanu, Dan S. and Brown, W. Michael and Crozier, Paul S. and In 'T Veld, Pieter J. and Kohlmeyer, Axel and Moore, Stan G. and Nguyen, Trung Dac and Shan, Ray and Stevens, Mark J. and Tranchida, Julien and Trott, Christian and Plimpton, Steven J.},
	month = feb,
	year = {2022},
	pages = {108171},
}

@article{ASE,
	title = {The atomic simulation environment—a {Python} library for working with atoms},
	volume = {29},
	issn = {0953-8984, 1361-648X},
	url = {https://iopscience.iop.org/article/10.1088/1361-648X/aa680e},
	doi = {10.1088/1361-648X/aa680e},
	number = {27},
	urldate = {2025-05-16},
	journal = {Journal of Physics: Condensed Matter},
	author = {Hjorth Larsen, Ask and Jørgen Mortensen, Jens and Blomqvist, Jakob and Castelli, Ivano E and Christensen, Rune and Dułak, Marcin and Friis, Jesper and Groves, Michael N and Hammer, Bjørk and Hargus, Cory and Hermes, Eric D and Jennings, Paul C and Bjerre Jensen, Peter and Kermode, James and Kitchin, John R and Leonhard Kolsbjerg, Esben and Kubal, Joseph and Kaasbjerg, Kristen and Lysgaard, Steen and Bergmann Maronsson, Jón and Maxson, Tristan and Olsen, Thomas and Pastewka, Lars and Peterson, Andrew and Rostgaard, Carsten and Schiøtz, Jakob and Schütt, Ole and Strange, Mikkel and Thygesen, Kristian S and Vegge, Tejs and Vilhelmsen, Lasse and Walter, Michael and Zeng, Zhenhua and Jacobsen, Karsten W},
	month = jul,
	year = {2017},
	pages = {273002},
}

@article{lammps_gpu_1,
	title = {Implementing molecular dynamics on hybrid high performance computers – short range forces},
	volume = {182},
	copyright = {https://www.elsevier.com/tdm/userlicense/1.0/},
	issn = {00104655},
	url = {https://linkinghub.elsevier.com/retrieve/pii/S0010465510005102},
	doi = {10.1016/j.cpc.2010.12.021},
	language = {en},
	number = {4},
	urldate = {2025-05-16},
	journal = {Computer Physics Communications},
	author = {Brown, W. Michael and Wang, Peng and Plimpton, Steven J. and Tharrington, Arnold N.},
	month = apr,
	year = {2011},
	pages = {898--911},
}

@article{lammps_gpu_2,
	title = {Implementing molecular dynamics on hybrid high performance computers – {Particle}–particle particle-mesh},
	volume = {183},
	copyright = {https://www.elsevier.com/tdm/userlicense/1.0/},
	issn = {00104655},
	url = {https://linkinghub.elsevier.com/retrieve/pii/S0010465511003444},
	doi = {10.1016/j.cpc.2011.10.012},
	language = {en},
	number = {3},
	urldate = {2025-05-16},
	journal = {Computer Physics Communications},
	author = {Brown, W. Michael and Kohlmeyer, Axel and Plimpton, Steven J. and Tharrington, Arnold N.},
	month = mar,
	year = {2012},
	pages = {449--459},
}

@article{lammps_gpu_3,
	title = {Implementing molecular dynamics on hybrid high performance computers—{Three}-body potentials},
	volume = {184},
	issn = {00104655},
	url = {https://linkinghub.elsevier.com/retrieve/pii/S0010465513002634},
	doi = {10.1016/j.cpc.2013.08.002},
	language = {en},
	number = {12},
	urldate = {2025-05-16},
	journal = {Computer Physics Communications},
	author = {Brown, W. Michael and Yamada, Masako},
	month = dec,
	year = {2013},
	pages = {2785--2793},
}

@article{lammps_gpu_4,
	title = {Accelerating dissipative particle dynamics simulations for soft matter systems},
	volume = {100},
	issn = {09270256},
	url = {https://linkinghub.elsevier.com/retrieve/pii/S0927025614007678},
	doi = {10.1016/j.commatsci.2014.10.068},
	language = {en},
	urldate = {2025-05-16},
	journal = {Computational Materials Science},
	author = {Nguyen, Trung Dac and Plimpton, Steven J.},
	month = apr,
	year = {2015},
	pages = {173--180},
}

@article{lammps_gpu_5,
	title = {{GPU}-accelerated {Tersoff} potentials for massively parallel {Molecular} {Dynamics} simulations},
	volume = {212},
	issn = {00104655},
	url = {https://linkinghub.elsevier.com/retrieve/pii/S0010465516303393},
	doi = {10.1016/j.cpc.2016.10.020},
	language = {en},
	urldate = {2025-05-16},
	journal = {Computer Physics Communications},
	author = {Nguyen, Trung Dac},
	month = mar,
	year = {2017},
	pages = {113--122},
}

@incollection{lammps_gpu_6,
	title = {{GPU} {Acceleration} of {Four}-{Site} {Water} {Models} in {LAMMPS}},
	copyright = {https://creativecommons.org/licenses/by-nc/4.0/},
	url = {https://www.medra.org/servlet/aliasResolver?alias=iospressISBN&isbn=978-1-64368-070-5&spage=565&doi=10.3233/APC200086},
	urldate = {2025-05-16},
	booktitle = {Advances in {Parallel} {Computing}},
	publisher = {IOS Press},
	author = {{Nikolskiy Vsevolod} and {Stegailov Vladimir}},
	year = {2020},
	doi = {10.3233/APC200086},
}

@article{VASP,
	title = {Efficient iterative schemes for \textit{ab initio} total-energy calculations using a plane-wave basis set},
	volume = {54},
	copyright = {http://link.aps.org/licenses/aps-default-license},
	issn = {0163-1829, 1095-3795},
	url = {https://link.aps.org/doi/10.1103/PhysRevB.54.11169},
	doi = {10.1103/PhysRevB.54.11169},
	language = {en},
	number = {16},
	urldate = {2025-05-17},
	journal = {Physical Review B},
	author = {Kresse, G. and Furthmüller, J.},
	month = oct,
	year = {1996},
	pages = {11169--11186},
}

@article{GGA_PBE,
	title = {Generalized {Gradient} {Approximation} {Made} {Simple}},
	volume = {77},
	copyright = {http://link.aps.org/licenses/aps-default-license},
	issn = {0031-9007, 1079-7114},
	url = {https://link.aps.org/doi/10.1103/PhysRevLett.77.3865},
	doi = {10.1103/PhysRevLett.77.3865},
	language = {en},
	number = {18},
	urldate = {2025-05-17},
	journal = {Physical Review Letters},
	author = {Perdew, John P. and Burke, Kieron and Ernzerhof, Matthias},
	month = oct,
	year = {1996},
	pages = {3865--3868},
}

@article{VASP2,
	title = {From ultrasoft pseudopotentials to the projector augmented-wave method},
	volume = {59},
	copyright = {http://link.aps.org/licenses/aps-default-license},
	issn = {0163-1829, 1095-3795},
	url = {https://link.aps.org/doi/10.1103/PhysRevB.59.1758},
	doi = {10.1103/PhysRevB.59.1758},
	language = {en},
	number = {3},
	urldate = {2025-05-17},
	journal = {Physical Review B},
	author = {Kresse, G. and Joubert, D.},
	month = jan,
	year = {1999},
	pages = {1758--1775},
}

@book{evans_statistical_2007,
	edition = {1st},
	title = {Statistical {Mechanics} of {Nonequilibrium} {Liquids}},
	isbn = {978-1-921313-22-6},
	url = {https://press.anu.edu.au/publications/statistical-mechanics-nonequilibrium-liquids},
	urldate = {2025-05-17},
	publisher = {ANU Press},
	author = {Evans, Denis J. and Morriss, Gary P.},
	month = aug,
	year = {2007},
	doi = {10.22459/smnl.08.2007},
}

@article{baktash_diffusion_2020,
	title = {Diffusion of lithium ions in {Lithium}-argyrodite solid-state electrolytes},
	volume = {6},
	copyright = {https://creativecommons.org/licenses/by/4.0},
	issn = {2057-3960},
	url = {https://www.nature.com/articles/s41524-020-00432-1},
	doi = {10.1038/s41524-020-00432-1},
	language = {en},
	number = {1},
	urldate = {2025-05-17},
	journal = {npj Computational Materials},
	author = {Baktash, Ardeshir and Reid, James C. and Roman, Tanglaw and Searles, Debra J.},
	month = oct,
	year = {2020},
	note = {Publisher: Springer Science and Business Media LLC},
}

@article{ong_phase_2013,
	title = {Phase stability, electrochemical stability and ionic conductivity of the {Li}$_{\textrm{10±1}}${MP}$_{\textrm{2}}${X}$_{\textrm{12}}$({M} = {Ge}, {Si}, {Sn}, {Al} or {P}, and {X} = {O}, {S} or {Se}) family of superionic conductors},
	volume = {6},
	issn = {1754-5692, 1754-5706},
	url = {https://xlink.rsc.org/?DOI=C2EE23355J},
	doi = {10.1039/c2ee23355j},
	language = {en},
	number = {1},
	urldate = {2025-05-17},
	journal = {Energy Environ. Sci.},
	author = {Ong, Shyue Ping and Mo, Yifei and Richards, William Davidson and Miara, Lincoln and Lee, Hyo Sug and Ceder, Gerbrand},
	year = {2013},
	note = {Publisher: Royal Society of Chemistry (RSC)},
	pages = {148--156},
}

@article{xu_machine_2023,
	title = {Machine learning molecular dynamics simulation identifying weakly negative effect of polyanion rotation on {Li}-ion migration},
	volume = {9},
	copyright = {https://creativecommons.org/licenses/by/4.0},
	issn = {2057-3960},
	url = {https://www.nature.com/articles/s41524-023-01049-w},
	doi = {10.1038/s41524-023-01049-w},
	language = {en},
	number = {1},
	urldate = {2025-05-17},
	journal = {npj Computational Materials},
	author = {Xu, Zhenming and Duan, Huiyu and Dou, Zhi and Zheng, Mingbo and Lin, Yixi and Xia, Yinghui and Zhao, Haitao and Xia, Yongyao},
	month = jun,
	year = {2023},
	note = {Publisher: Springer Science and Business Media LLC},
}

@article{LYB_LYC_exp,
	title = {Insights into the {Lithium} {Sub}-structure of {Superionic} {Conductors} {Li3YCl6} and {Li3YBr6}},
	volume = {33},
	issn = {0897-4756},
	url = {https://doi.org/10.1021/acs.chemmater.0c04352},
	doi = {10.1021/acs.chemmater.0c04352},
	number = {1},
	journal = {Chemistry of Materials},
	author = {Schlem, Roman and Banik, Ananya and Ohno, Saneyuki and Suard, Emmanuelle and Zeier, Wolfgang G.},
	month = jan,
	year = {2021},
	note = {Publisher: American Chemical Society},
	pages = {327--337},
}

@article{7net,
	title = {Scalable Parallel Algorithm for Graph Neural Network Interatomic Potentials in Molecular Dynamics Simulations},
	volume = {20},
	doi = {10.1021/acs.jctc.4c00190},
	number = {11},
	journal = {J. Chem. Theory Comput.},
	author = {Park, Yutack and Kim, Jaesun and Hwang, Seungwoo and Han, Seungwu},
	year = {2024},
	pages = {4857--4868},
}

@article{7net_mf_ompa,
	title = {Data-Efficient Multifidelity Training for High-Fidelity Machine Learning Interatomic Potentials},
	volume = {147},
	doi = {10.1021/jacs.4c14455},
	number = {1},
	journal = {J. Am. Chem. Soc.},
	author = {Kim, Jaesun and Kim, Jisu and Kim, Jaehoon and Lee, Jiho and Park, Yutack and Kang, Youngho and Han, Seungwu},
	year = {2024},
	pages = {1042--1054},
}

@misc{park_sevennet_mf_ompa_2025,
	title = {{SevenNet}\_MF\_ompa},
	copyright = {GPL 3.0+},
	url = {https://figshare.com/articles/software/7net_MF_ompa/28590722/7},
	urldate = {2025-05-20},
	publisher = {figshare},
	author = {Park, Yutack and Kim, Jaesun and Kim, Jisu and Jeon, Haekwan and Han, Seungwu},
	year = {2025},
	doi = {10.6084/M9.FIGSHARE.28590722.V7},
	note = {Artwork Size: 412242578 Bytes
Pages: 412242578 Bytes},
}

@article{ohno_materials_2020,
	title = {Materials design of ionic conductors for solid state batteries},
	volume = {2},
	issn = {2516-1083},
	url = {https://iopscience.iop.org/article/10.1088/2516-1083/ab73dd},
	doi = {10.1088/2516-1083/ab73dd},
	number = {2},
	urldate = {2025-05-31},
	journal = {Progress in Energy},
	author = {Ohno, Saneyuki and Banik, Ananya and Dewald, Georg F and Kraft, Marvin A and Krauskopf, Thorben and Minafra, Nicoló and Till, Paul and Weiss, Manuel and Zeier, Wolfgang G},
	month = mar,
	year = {2020},
	pages = {022001},
}

@article{daniel_materials_2008,
	title = {Materials and processing for lithium-ion batteries},
	volume = {60},
	copyright = {http://www.springer.com/tdm},
	issn = {1047-4838, 1543-1851},
	url = {http://link.springer.com/10.1007/s11837-008-0116-x},
	doi = {10.1007/s11837-008-0116-x},
	language = {en},
	number = {9},
	urldate = {2025-06-01},
	journal = {JOM},
	author = {Daniel, Claus},
	month = sep,
	year = {2008},
	pages = {43--48},
}

@article{goodenough_li-ion_2013,
	title = {The {Li}-{Ion} {Rechargeable} {Battery}: {A} {Perspective}},
	volume = {135},
	issn = {0002-7863, 1520-5126},
	shorttitle = {The {Li}-{Ion} {Rechargeable} {Battery}},
	url = {https://pubs.acs.org/doi/10.1021/ja3091438},
	doi = {10.1021/ja3091438},
	language = {en},
	number = {4},
	urldate = {2025-06-01},
	journal = {Journal of the American Chemical Society},
	author = {Goodenough, John B. and Park, Kyu-Sung},
	month = jan,
	year = {2013},
	pages = {1167--1176},
}

@article{janek_challenges_2023,
	title = {Challenges in speeding up solid-state battery development},
	volume = {8},
	issn = {2058-7546},
	url = {https://www.nature.com/articles/s41560-023-01208-9},
	doi = {10.1038/s41560-023-01208-9},
	language = {en},
	number = {3},
	urldate = {2025-06-01},
	journal = {Nature Energy},
	author = {Janek, Jürgen and Zeier, Wolfgang G.},
	month = feb,
	year = {2023},
	pages = {230--240},
}

@article{kerman2017review,
  title={Review—Practical challenges hindering the development of solid-state Li-ion batteries},
  author={Kerman, K. and Luntz, A. and Viswanathan, V. and Chiang, Y.-M. and Chen, Z.},
  journal={Journal of The Electrochemical Society},
  volume={164},
  number={7},
  pages={A1731--A1744},
  year={2017},
  doi={10.1149/2.1571707jes}
}

@article{pasta2020roadmap,
  title={Energy 2020 roadmap on solid-state batteries},
  author={Pasta, Mauro and others},
  journal={Journal of Physics: Energy},
  volume={2},
  number={3},
  pages={032008},
  year={2020},
  doi={10.1088/2515-7655/ab8fce}
}

@article{asano_solid_2018,
	title = {Solid {Halide} {Electrolytes} with {High} {Lithium}‐{Ion} {Conductivity} for {Application} in 4 {V} {Class} {Bulk}‐{Type} {All}‐{Solid}‐{State} {Batteries}},
	volume = {30},
	issn = {0935-9648, 1521-4095},
	url = {https://onlinelibrary.wiley.com/doi/10.1002/adma.201803075},
	doi = {10.1002/adma.201803075},
	language = {en},
	number = {44},
	urldate = {2025-06-01},
	journal = {Advanced Materials},
	author = {Asano, Tetsuya and Sakai, Akihiro and Ouchi, Satoru and Sakaida, Masashi and Miyazaki, Akinobu and Hasegawa, Shinya},
	month = nov,
	year = {2018},
	pages = {1803075},
}

@article{van_der_maas_investigation_2023,
	title = {Investigation of {Structure}, {Ionic} {Conductivity}, and {Electrochemical} {Stability} of {Halogen} {Substitution} in {Solid}-{State} {Ion} {Conductor} {Li3YBrxCl6}–x},
	volume = {127},
	issn = {1932-7447},
	url = {https://doi.org/10.1021/acs.jpcc.2c07910},
	doi = {10.1021/acs.jpcc.2c07910},
	number = {1},
	journal = {The Journal of Physical Chemistry C},
	author = {van der Maas, Eveline and Zhao, Wenxuan and Cheng, Zhu and Famprikis, Theodosios and Thijs, Michel and Parnell, Steven R. and Ganapathy, Swapna and Wagemaker, Marnix},
	month = jan,
	year = {2023},
	note = {Publisher: American Chemical Society},
	pages = {125--132},
}

@article{liu_tuning_2024,
	title = {Tuning collective anion motion enables superionic conductivity in solid-state halide electrolytes},
	volume = {16},
	issn = {1755-4330, 1755-4349},
	url = {https://www.nature.com/articles/s41557-024-01634-6},
	doi = {10.1038/s41557-024-01634-6},
	language = {en},
	number = {10},
	urldate = {2025-06-04},
	journal = {Nature Chemistry},
	author = {Liu, Zhantao and Chien, Po-Hsiu and Wang, Shuo and Song, Shaowei and Lu, Mu and Chen, Shuo and Xia, Shuman and Liu, Jue and Mo, Yifei and Chen, Hailong},
	month = oct,
	year = {2024},
	pages = {1584--1591},
}

@article{lejaeghere_reproducibility_2016,
	title = {Reproducibility in density functional theory calculations of solids},
	volume = {351},
	issn = {0036-8075, 1095-9203},
	url = {https://www.science.org/doi/10.1126/science.aad3000},
	doi = {10.1126/science.aad3000},
	language = {en},
	number = {6280},
	urldate = {2025-06-06},
	journal = {Science},
	author = {Lejaeghere, Kurt and Bihlmayer, Gustav and Björkman, Torbjörn and Blaha, Peter and Blügel, Stefan and Blum, Volker and Caliste, Damien and Castelli, Ivano E. and Clark, Stewart J. and Dal Corso, Andrea and De Gironcoli, Stefano and Deutsch, Thierry and Dewhurst, John Kay and Di Marco, Igor and Draxl, Claudia and Dułak, Marcin and Eriksson, Olle and Flores-Livas, José A. and Garrity, Kevin F. and Genovese, Luigi and Giannozzi, Paolo and Giantomassi, Matteo and Goedecker, Stefan and Gonze, Xavier and Grånäs, Oscar and Gross, E. K. U. and Gulans, Andris and Gygi, François and Hamann, D. R. and Hasnip, Phil J. and Holzwarth, N. A. W. and Iuşan, Diana and Jochym, Dominik B. and Jollet, François and Jones, Daniel and Kresse, Georg and Koepernik, Klaus and Küçükbenli, Emine and Kvashnin, Yaroslav O. and Locht, Inka L. M. and Lubeck, Sven and Marsman, Martijn and Marzari, Nicola and Nitzsche, Ulrike and Nordström, Lars and Ozaki, Taisuke and Paulatto, Lorenzo and Pickard, Chris J. and Poelmans, Ward and Probert, Matt I. J. and Refson, Keith and Richter, Manuel and Rignanese, Gian-Marco and Saha, Santanu and Scheffler, Matthias and Schlipf, Martin and Schwarz, Karlheinz and Sharma, Sangeeta and Tavazza, Francesca and Thunström, Patrik and Tkatchenko, Alexandre and Torrent, Marc and Vanderbilt, David and Van Setten, Michiel J. and Van Speybroeck, Veronique and Wills, John M. and Yates, Jonathan R. and Zhang, Guo-Xu and Cottenier, Stefaan},
	month = mar,
	year = {2016},
	pages = {aad3000},
}

@article{wang_lithium_2019,
	title = {Lithium {Chlorides} and {Bromides} as {Promising} {Solid}‐{State} {Chemistries} for {Fast} {Ion} {Conductors} with {Good} {Electrochemical} {Stability}},
	volume = {58},
	issn = {1433-7851, 1521-3773},
	url = {https://onlinelibrary.wiley.com/doi/10.1002/anie.201901938},
	doi = {10.1002/anie.201901938},
	language = {en},
	number = {24},
	urldate = {2025-06-09},
	journal = {Angewandte Chemie International Edition},
	author = {Wang, Shuo and Bai, Qiang and Nolan, Adelaide M. and Liu, Yunsheng and Gong, Sheng and Sun, Qiang and Mo, Yifei},
	month = jun,
	year = {2019},
	pages = {8039--8043},
}

@article{he_statistical_2018,
	title = {Statistical variances of diffusional properties from ab initio molecular dynamics simulations},
	volume = {4},
	issn = {2057-3960},
	url = {https://www.nature.com/articles/s41524-018-0074-y},
	doi = {10.1038/s41524-018-0074-y},
	language = {en},
	number = {1},
	urldate = {2025-06-10},
	journal = {npj Computational Materials},
	author = {He, Xingfeng and Zhu, Yizhou and Epstein, Alexander and Mo, Yifei},
	month = apr,
	year = {2018},
	pages = {18},
}

@article{devita_on_the_fly,
  title = {Molecular Dynamics with On-the-Fly Machine Learning of Quantum-Mechanical Forces},
  author = {Li, Zhenwei and Kermode, James R. and De Vita, Alessandro},
  journal = {Phys. Rev. Lett.},
  volume = {114},
  issue = {9},
  pages = {096405},
  numpages = {5},
  year = {2015},
  month = {Mar},
  publisher = {American Physical Society},
  doi = {10.1103/PhysRevLett.114.096405},
  url = {https://link.aps.org/doi/10.1103/PhysRevLett.114.096405}
}

@article{shapeev_active_learning,
author = {Shapeev, Alexander V.},
title = {Moment Tensor Potentials: A Class of Systematically Improvable Interatomic Potentials},
journal = {Multiscale Modeling \& Simulation},
volume = {14},
number = {3},
pages = {1153-1173},
year = {2016},
doi = {10.1137/15M1054183},
URL = {https://doi.org/10.1137/15M1054183},
eprint = {https://doi.org/10.1137/15M1054183}
}

@article{bernstein_active_learning,
author = {Bernstein, Noam and Csányi, Gábor and Deringer, Volker L.},
title = {De novo exploration and self-guided learning of potential-energy surfaces},
journal = {npj Computational Materials},
volume = {5},
number = {1},
pages = {99},
year = {2019},
doi = {10.1038/s41524-019-0236-6},
URL = {https://doi.org/10.1038/s41524-019-0236-6},
eprint = {https://doi.org/10.1038/s41524-019-0236-6}
}

@article{vandermause_on_the_fly,
author = {Vandermause, Jonathan and Torrisi, Steven B. and Batzner, Simon and Xie, Yu and Sun, Lixin and Kolpak, Alexie M. and Kozinsky, Boris},
title = {On-the-fly active learning of interpretable Bayesian force fields for atomistic rare events},
journal = {npj Computational Materials},
volume = {6},
number = {1},
pages = {20},
year = {2020},
doi = {10.1038/s41524-020-0283-z},
URL = {https://doi.org/10.1038/s41524-020-0283-z},
eprint = {https://doi.org/10.1038/s41524-020-0283-z}
}
